\documentclass[fleqn,usenatbib]{mnras}

\usepackage[T1]{fontenc}
\usepackage{ae,aecompl}

\usepackage{graphicx}	
\usepackage{amsmath}	
\usepackage{amssymb}	
\usepackage{multirow}
\usepackage{url}
\usepackage{footmisc}
\usepackage{enumitem}
\usepackage{caption}
\usepackage{hyperref}
\usepackage{mathptmx}
\usepackage{txfonts}
\usepackage{pdfpages} 
\usepackage{pgffor} 

\makeatletter
\AtBeginDocument{\let\LS@rot\@undefined}
\makeatother

\def\supplementfilename{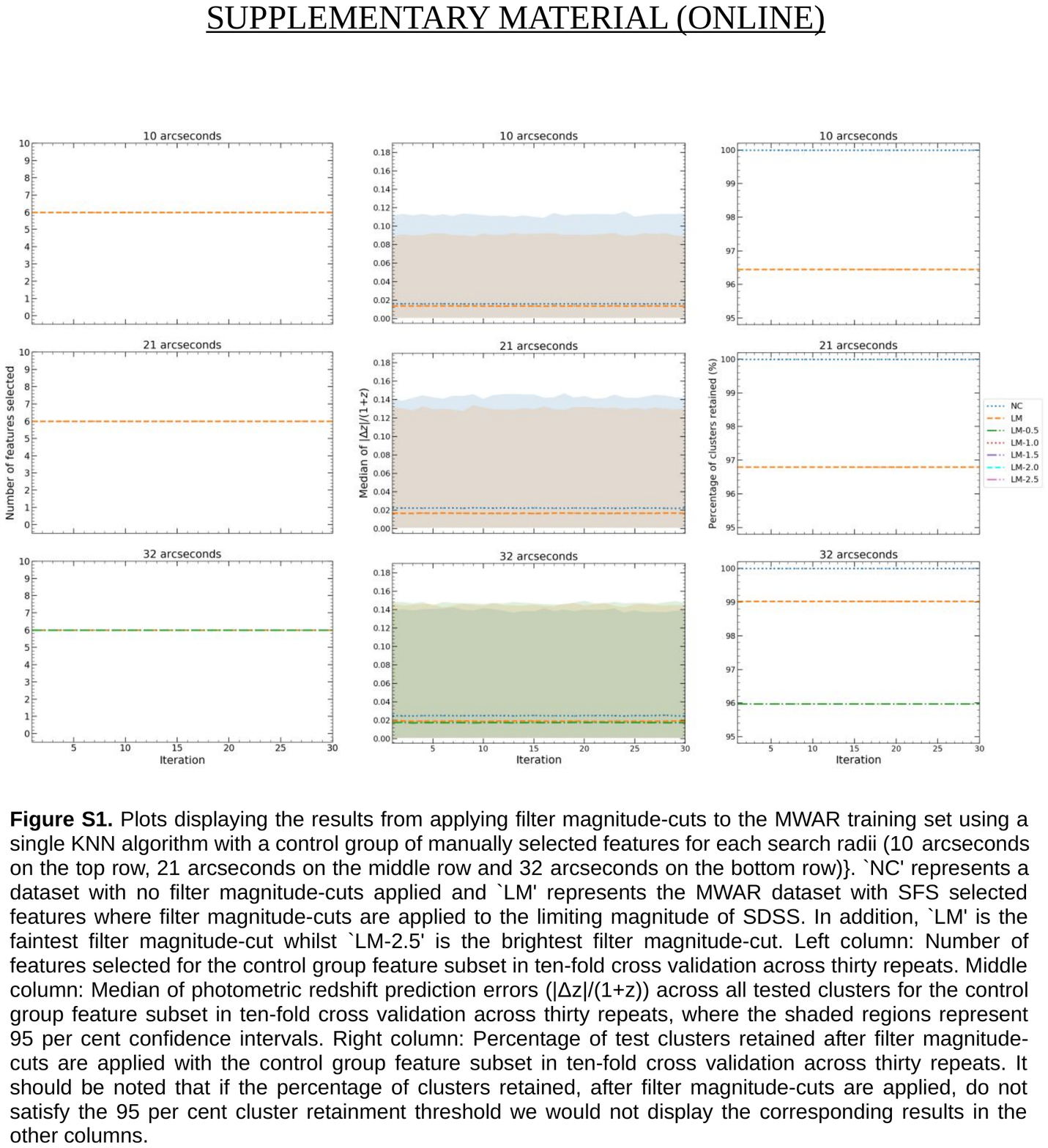}

\pdfximage{\supplementfilename}
\def\numbersupplementpages{\the\pdflastximagepages}

\newif\ifarXiv
\arXivtrue 

\title[Estimating Photometric Redshifts Of Clusters]{Z-Sequence: Photometric redshift predictions for galaxy clusters with sequential random k-nearest neighbours}

\author[M. C. Chan and J. P. Stott]{
Matthew C. Chan$^{1}$ and John P. Stott$^{1}$
\\
\texttt{E-mails: m.c.chan@lancaster.ac.uk and j.p.stott@lancaster.ac.uk}
\\
$^{1}$Department of Physics, Lancaster University, Lancaster, LA1 4YB, UK
}

\date{Accepted XXX. Received YYY; in original form ZZZ}

\pubyear{2021}

\begin{document}
\label{firstpage}
\pagerange{\pageref{firstpage}--\pageref{lastpage}}
\maketitle

\begin{abstract}
\label{abstract}

We introduce Z-Sequence, a novel empirical model that utilises photometric measurements of observed galaxies within a specified search radius to estimate the photometric redshift of galaxy clusters. Z-Sequence itself is composed of a machine learning ensemble based on the k-nearest neighbours algorithm. We implement an automated feature selection strategy that iteratively determines appropriate combinations of filters and colours to minimise photometric redshift prediction error. We intend for Z-Sequence to be a standalone technique but it can be combined with cluster finders that do not intrinsically predict redshift, such as our own DEEP-CEE. In this proof-of-concept study we train, fine-tune and test Z-Sequence on publicly available cluster catalogues derived from the Sloan Digital Sky Survey. We determine the photometric redshift prediction error of Z-Sequence via the median value of $|\Delta z|/(1+z)$ (across a photometric redshift range of $0.05 \le \textit{z} \le 0.6$) to be $\sim0.01$ when applying a small search radius. The photometric redshift prediction error for test samples increases by 30-50 per cent when the search radius is enlarged, likely due to line-of-sight interloping galaxies. Eventually, we aim to apply Z-Sequence to upcoming imaging surveys such as the Legacy Survey of Space and Time to provide photometric redshift estimates for large samples of as yet undiscovered and distant clusters.

\end{abstract}




\section{Introduction}
\label{introduction}

Galaxy clusters are the most massive gravitationally bound objects to have formed in the Universe, with deep potential wells that correspond to matter density peaks (\citealt{cluster_formation_0}; \citealt{cluster_formation_1}; \citealt{cluster_formation_2}). During the past few decades, the advent of modern imaging surveys has significantly contributed to the study of large scale structure and galaxy evolution across cosmic time. These surveys generate a huge abundance of data that encourages the need for automated algorithms (e.g. \citealt{ssrs_redshift_survey}; \citealt{cfa2_redshift_survey}; \citealt{sdss_technical_0}; \citealt{2df_redshift_survey}; \citealt{6df_redshift_survey}; \citealt{gama_redshift_survey}; \citealt{sdss_III}; \citealt{2mass_redshift_survey}; \citealt{sdss_iv}). From which, properties of clusters such as redshift, luminosity and richness can be estimated and used as probes for astrophysics and cosmology (e.g. \citealt{cluster_probe_0}; \citealt{cluster_probe_1}; \citealt{cluster_probe_2}; \citealt{cluster_probe_3}; \citealt{cluster_probe_4}; \citealt{cluster_probe_5}; \citealt{cluster_probe_6}; \citealt{cluster_probe_7}; \citealt{cluster_probe_8}; \citealt{cluster_probe_9}; \citealt{cluster_probe_10}).  

The Legacy Survey of Space and Time (LSST\footnote[1]{LSST will be conducted using the 8.4-meter Simonyi Survey Telescope at the Vera Rubin Observatory operating with six broad-band filters: \textit{u, g, r, i, z and Y}.}, \citealt{lsst_survey}) will be the state-of-the-art imaging survey for the next decade of astronomy. It will repeatedly image the entire southern hemisphere and is forecasted to generate up to twenty terabytes of data per night over a ten year period. Due to the quantity of data involved, the development of automated algorithms for LSST will be crucial to handle extensive data processing and analysis tasks. In addition, LSST will observe at deeper depths and wider sky coverage compared to previous surveys. This would increase the redshift range and lower the mass limit sensitivity of current cluster observations, such that thousands of new clusters are likely to be discovered.

There are presently two approaches used to determine galaxy redshifts, these are through spectroscopy and photometry (e.g. \citealt{redshift_measurements_0}; \citealt{redshift_measurements_1}). However whilst the former is precise it is also time-consuming, expensive and difficult to perform for faint distant sources, which limits the number of observations with spectroscopic redshifts. Alternatively, photometric redshifts are fast to acquire and have been shown to be successful for faint distant sources (e.g. \citealt{photometry_faint_distant_sources}). Conventional methods to estimate photometric redshift involve either empirical or template fitting algorithms. Empirical algorithms learn a target function of the underlying relationships between observed brightness, colour and spectroscopic redshift from a large training sample of galaxies (e.g. \citealt{empirical_algorithm_0}; \citealt{empirical_algorithm_1}; \citealt{empirical_algorithm_2}; \citealt{empirical_algorithm_3}; \citealt{empirical_algorithm_4}). Whilst, template fitting algorithms match observed fluxes to theoretical spectral energy distributions of different galaxy types at reference redshifts (e.g. \citealt{template_fitting_0}; \citealt{template_fitting_1}; \citealt{template_fitting_2}; \citealt{template_fitting_3}). Nevertheless, photometric redshifts tend to have larger measurement errors than spectroscopic redshifts since photometric filters operate with low wavelength resolution, which means that individual spectral features can not be utilised to determine redshift. 

Photometric redshifts are often employed by imaging surveys to provide initial redshift estimates for many galaxies (e.g. \citealt{photometric_redshift_survey_0}; \citealt{photometric_redshift_survey_1}; \citealt{photometric_redshift_survey_2}; \citealt{photometric_redshift_survey_3}), of which sub-samples can be followed up with spectroscopic redshifts. Similarly, it is important to develop models that will provide researchers with accurate initial redshift estimates for large and deep samples of the cluster population. In terms of predictive power for the low to intermediate redshift regime, empirical algorithms with sufficient training samples will generally outperform template fitting algorithms because template fitting algorithms require more physical assumptions when constructing spectral energy distributions to reflect possible observations. Whereas for the high redshift regime, template fitting algorithms will typically outperform empirical algorithms since high redshift training samples are more difficult to obtain due to observing limitations \citep{photometric_redshift_empirical_vs_template}.

In order to estimate redshifts for clusters, it is first required to identify cluster members within a given search area. This can be conducted by utilising the red sequence (\citealt{red_sequence_detection_0}; \citealt{red_sequence_detection_1}), which takes advantage of the fact that `red' early-type galaxies are often found in clusters \citep{cluster_population}. From which, the red sequence is seen as a well-defined linear relationship in colour-magnitude space (CMS) that evolves with redshift \citep{red_sequence_evolution}. This sequence is sloped such that bright cluster members are redder than their fainter counterparts. In CMS, galaxy types can be differentiated based on their underlying stellar populations into a red sequence and blue cloud region \citep{color_magnitude_diagrams_separation}. Generally, the red sequence contains predominately `red' elliptical and lenticular galaxies, whilst the blue cloud contains mostly `blue' spiral and `disk'-like galaxies. However, minority exceptions do exist such as `red' spiral galaxies \citep{galaxy_types_red_spirals} and `blue' elliptical galaxies \citep{galaxy_types_blue_ellipticals}. From which, an empirical algorithm can estimate photometric redshift based on the observed red sequence (e.g. \citealt{cluster_redshift_prediction_0}; \citealt{redMapper}). This involves training an empirical algorithm to learn the redshifts from examples of known red sequences, such that the red sequence of an unknown cluster can be interpolated by the algorithm.

Additionally in order to break any colour-redshift degeneracies, where galaxies at different redshifts could have resembling colours, multi-dimensional CMS should be employed to reduce the reliance on specific colours. For example, a single colour that only utilised short wavelength optical filters would struggle to detect the red sequence of a high redshift cluster since the filters would be unable to observe the redshifted 4000\AA \ break\footnote[2]{The 4000\AA \ break is caused by the blanket absorption of photons at specific wavelengths from metals in the ionised atmospheres of old stellar populations \citep{4000_angstrom_break}.}, which is a distinctive broad spectral feature seen in the continuum spectrum of elliptical galaxies \citep{4000_angstrom_break_ellipticals}. By utilising more colours, it is possible to straddle the 4000\AA \ break to account for its transition at different redshifts (\citealt{red_sequence_detection_1}; \citealt{red_sequence_detection}).

For this paper, we employ an automated feature selection strategy that selects appropriate combinations of filters and colours in multi-dimensional CMS. We intend for this feature selection process to be fully data-driven based on observed galaxy photometry data, such that the selected features are effective at minimising photometric redshift prediction error. This method also comes with multiple practical benefits. Firstly, it is able to work with incomplete filter sets, as it does not rely on any specific filter. Secondly, it does not depend on galaxy photometric redshift catalogues. Thirdly, this approach can be combined with cluster finders that do not naturally predict redshift, such as DEEP-CEE \citep{deep_cee}, since Z-Sequence only requires input astronomical coordinates and a photometry catalogue to predict photometric redshift of clusters.

We structure this paper with the following layout. In \S\ref{sec:methodology} we outline our methodology where \S\S\ref{sec:prepare_photometry_from_catalogues} describes our data pre-processing approach, \S\S\ref{sec:machine_learning_method} describes our feature selection strategy plus machine learning algorithm and \S\S\ref{sec:outline_of_approach} describes how we train our model. In \S\ref{sec:results} we present our results where \S\S\ref{sec:feature_selection_analysis} describes the feature selection and filter magnitude-cut analysis, \S\S\ref{sec:hyper-parameter_analysis} describes the hyper-parameter tuning and \S\S\ref{sec:Model_Analysis_with_Test_Set} plus \S\S\ref{sec:Model_Analysis_with_Additional_Test_Set} describes the tuned model performance on test sets. In \S\ref{sec:Discussion} we review our findings where \S\S\ref{sec:z-sequence_discussion} discusses the effectiveness of the tuned model at making predictions and \S\S\ref{sec:machine_learning_discussion} discusses the practicality of the machine learning techniques used in this paper. Finally, in \S\ref{sec:Conclusion} we summarise this paper. 

We assume the $\Lambda$CDM cosmological parameters $H_{0} = 71 \ \text{km} \ \text{s}^{-1} \ \text{Mpc}^{-1}$, $\Omega_{m} = 0.27$ and $\Omega_{\Lambda} = 0.73$.

\section{Methodology}
\label{sec:methodology}

\subsection{Preparation Of Photometric Datasets}
\label{sec:prepare_photometry_from_catalogues} 

We utilise candidate clusters detected in the Sloan Digital Sky Survey III (SDSS-III, \citealt{sdss_III}) by the WHL12 \citep{whl} and redMaPPer \citep{redMapper} cluster catalogues as part of our training, validation and test sets under a supervised learning approach \citep{supervised_learning}. WHL12 uses photometric redshifts of galaxies estimated by SDSS to identify overdense regions of galaxy clustering via a grouping algorithm, in which the cluster redshift was calculated from the median value of determined cluster members. Whilst redMaPPer search for the red sequence within CMS across the SDSS sky coverage. The observed red sequence profile of highly probable cluster members was then fit with a self-trained model of template red sequences to estimate cluster redshift. It should be noted that the full WHL12 cluster catalogue has a photometric redshift range of $0.05 \leq \textit{z} \leq 0.7846$ and the full redMaPPer cluster catalogue has a photometric redshift range of $0.0811 \leq \textit{z} \leq 0.5983$.
   
Initially, we apply two selection criterion to the WHL12 cluster catalogue to identify clusters that had photometric redshifts between $0.0 < \textit{z} < 0.6$ and also contain more than twenty member galaxies. This provides us with an approximation of the distribution of clusters found at different redshifts. From which, we calculate a mean photometric redshift of $\textit{z} = 0.3127$ based on the selected clusters. We use this mean photometric redshift to determine an angular distance of 54.96 arcseconds, which corresponds with the average cluster core optical radius of $\sim250$ kpc \citep{cluster_radius}. This angular distance also corresponds to a radius of approximately 100 kpc at $\textit{z} = 0.1$ and 334 kpc at $\textit{z} = 0.5$. We then cross-match the clusters from the full WHL12 and redMaPPer cluster catalogues that are within 54.96 arcseconds and also within a photometric redshift range of $\pm 0.04(1 + \textit{z})$ as used by \cite{photoz_gap}\footnote[3]{\cite{photoz_gap} suggests that a photometric redshift gap of $\pm 0.04(1 + \textit{z})$ is a suitable indicator of true cluster richness, which corresponds to a rest frame velocity range of 24000 km $\text{s}^{-1}$ to account for the uncertainty of the photometric redshifts.}. This ensures that we cleanly separate clusters to improve signal-to-noise in the dataset. The matching and non-matching clusters are then split into the following three datasets: 

\begin{itemize}[leftmargin=0.3cm]
\item \textbf{MWAR} - Cross-\textbf{m}atched \textbf{W}HL12 \textbf{a}nd \textbf{r}edMapper clusters.
\item \textbf{WNMR} - \textbf{W}HL12 clusters with \textbf{n}o cross-\textbf{m}atched \textbf{r}edMapper clusters.
\item \textbf{RNMW} - \textbf{r}edMapper clusters with \textbf{n}o cross-\textbf{m}atched \textbf{W}HL12 clusters.
\end{itemize}

\begin{figure}
\begin{center}
	\includegraphics[width=0.95\linewidth, height=0.85\linewidth]{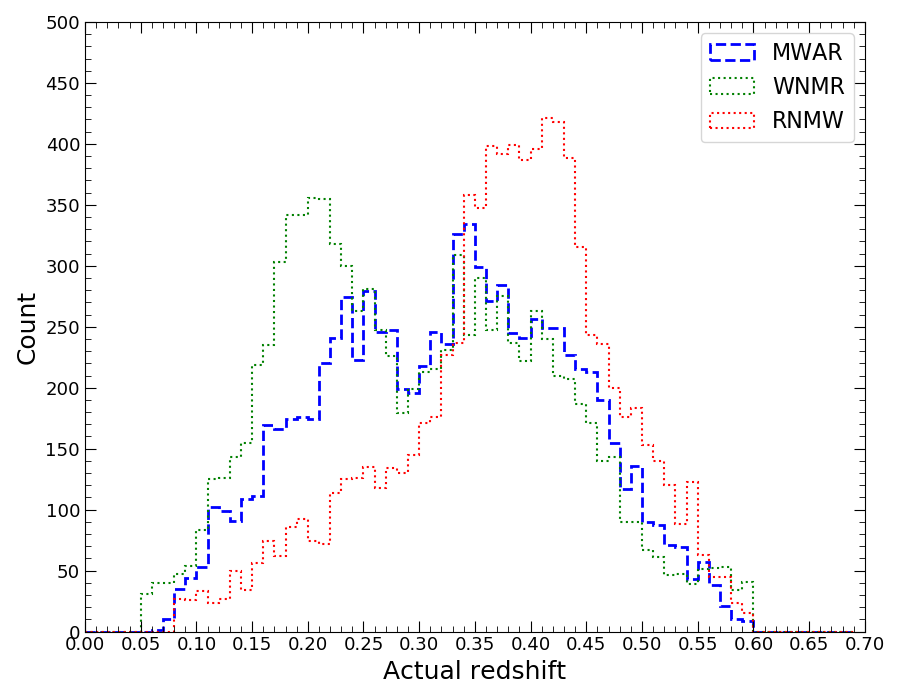}
    \caption{Frequency histogram of the `actual' redshift distributions of clusters, where photometric redshifts of clusters in the MWAR (blue dashed line) and WNMR (green dotted line) datasets are originally estimated by WHL12. Whilst the photometric redshifts of clusters in the RNMW (red dotted line) dataset are originally estimated by redMaPPer.}
    \label{fig:redshift_population}
\end{center}
\end{figure}

Next, we reapply our initial two selection criterion to all the clusters in the MWAR, RNMW and WNMR datasets. This splits the clusters in each dataset into distinctive redshift and richness groupings, which can be used to examine how the Z-Sequence model performs on clusters that have these different properties. We set clusters that have properties within the selection criterion limits as the main training and test sets, whilst clusters that have properties outside the selection criterion limits are used as additional test sets. From which, the number of clusters within the selection criterion limits for the MWAR dataset is 8841 with a photometric redshift range of $0.0698 \leq \textit{z} \leq 0.5986$, the WNMR dataset is 9723 with a photometric redshift range of $0.05 \leq \textit{z} \leq 0.599$ and the RNMW dataset is 8646 with a photometric redshift range of $0.0811 \leq \textit{z} \leq 0.5983$. In addition, the observed redshift distributions and positions of clusters from each dataset can be seen in Figures \ref{fig:redshift_population} and SA1 (available online).

We proceed to cross-match the astronomical coordinates of clusters in each dataset to galaxies found in the SDSS-III Data Release 9 photometric catalogue (SDSS-III DR9, \citealt{sdss_dr9_photometry}) that are within the previously defined angular distance of 54.96 arcseconds. We select `primary' observations\footnote[4]{The term `primary' refers to the best imaging observation recorded for a survey object if it was seen multiple times during an observing run in an SDSS plate, whilst other observations of the object are called `secondary'. A more in-depth explanation can be found on \url{http://www.sdss3.org/dr9/help/glossary.php}} of galaxies that have `clean' photometry as determined by SDSS. This catalogue provides photometric measurements\footnote[5]{SDSS `modelMag' measurements are used for filter magnitudes and colours of galaxies. This approach ensures the same aperture is used for all filters and the resultant magnitudes are calculated based off the best-fit model parameters observed in the r-band. For further details see \url{http://www.sdss3.org/dr9/algorithms/magnitudes.php}} for the following filters and colours:

\begin{itemize}[leftmargin=0.3cm]
\item \textbf{Filters}: \textit{u, g, r, i, z},
\item \textbf{Colours}: \textit{u-g, g-r, r-i, i-z , u-r, g-i, r-z, u-i, g-z, u-z},
\end{itemize}
where we use these filters and colours as our input features in \S\S\ref{sec:machine_learning_method}.

We assume that any of the SDSS identified galaxies which lie along the line-of-sight and within 54.96 arcseconds of the input astronomical coordinates are part of the same cluster, from which we assign each individual galaxy a cluster ID number for cross-referencing. To reduce the number of interloped galaxies, we empirically set multiple search radii of approximately 50, 100 and 150 kpc at the mean photometric redshift of $\textit{z} = 0.3127$, which corresponds to angular distances of 10, 21 and 32 arcseconds respectively. The number of interlopers will also depend on the position accuracy of the input cluster coordinates relative to the true cluster centroid. The reason we employ multiple search radii was to ensure that if the smallest search radius did not find a galaxy in the SDSS-III DR9 photometric catalogue, then the search radius would increase until a galaxy was found. This also provides a test for the effectiveness of the algorithm when given different views of the cluster core. It should be noted that this results in multiple forms of the training/validation/test sets that contain additional galaxies in clusters found within each search radius.

We assign the MWAR dataset as the training/validation sets and WNMR/RNMW datasets as test sets. The redshift distributions of the clusters in these datasets can be seen in Figure SA2 (available online) for each search radius. We chose the MWAR dataset as the training set since we expect that these clusters would be more likely to host a populated core, where the red sequence would be well-defined (\citealt{red_sequence_0}; \citealt{red_sequence_1}; \citealt{red_sequence_2}; \citealt{red_sequence_3}; \citealt{red_sequence_4}; \citealt{red_sequence_5}; \citealt{red_sequence_6}) in comparison to clusters in the WNMR/RNMW datasets, given the nature of the methods of WHL12 and redMaPPer. We want our model to learn and utilise `red sequence'-like features found within high dimensional CMS to effectively predict photometric redshifts across a broad redshift range.

Finally, we investigate how varying the brightness for filter magnitude-cuts (see Table \ref{tab:mag_cuts}) could improve the accuracy of photometric redshift estimates, as this will remove galaxies from the less well-defined faint end of the red sequence that have relatively large filter magnitude errors and filter magnitude values fainter than a specified limiting magnitude\footnote[6]{Limiting magnitudes for the SDSS telescope are found by repeated observations of a patch of sky to obtain a magnitude value that provides 95 per cent completeness of point sources (\citealt{sdss_technical_0}; \citealt{sdss_technical_1}; \citealt{sdss_technical_2}). See SDSS imaging camera scope at \url{http://www.sdss3.org/dr9/scope.php} for magnitude limits of each filter. \label{footnote_6}} value. In addition, we also compare the performance of using filter magnitude-cuts to a control group dataset that had no filter magnitude-cuts applied.

\begin{table}
	\begin{tabular}{|c|c|c|c|c|c|c|}
		\hline
	    Filter & LM & LM-0.5 & LM-1.0 & LM-1.5 & LM-2.0 & LM-2.5 \\
	     & [mag] & [mag] & [mag] & [mag] & [mag] & [mag] \\
		\hline
		\textit{u} & 21.6 & 21.1 & 20.6 & 20.1 & 19.6 & 19.1 \\
		\textit{g} & 22.2 & 21.7 & 21.2 & 20.7 & 20.2 & 19.7 \\
	    \textit{r} & 22.2 & 21.7 & 21.2 & 20.7 & 20.2 & 19.7 \\
	    \textit{i} & 21.3 & 20.8 & 20.3 & 19.8 & 19.3 & 18.8 \\
	    \textit{z} & 20.7 & 20.2 & 19.7 & 19.2 & 18.7 & 18.2 \\
		\hline
	\end{tabular}
	\caption{This table contains the SDSS limiting magnitude (LM) values of each filter with specified magnitude-cuts. The LM values are determined from 95 per cent completeness studies of point sources\textsuperscript{\ref{footnote_6}}. The filter magnitude values shown are converted from the SDSS \textit{ugriz} magnitude system \citep{sdss_ugriz_magnitude} to AB magnitude system \citep{ab_magnitude}. It should be noted that the SDSS \textit{ugriz} magnitude system is very similar to the AB magnitude system but not exact \citep{sdss_photometry_error}, such that $ u_{AB} = u_{SDSS} - 0.04$ and $z_{AB} = z_{SDSS} + 0.02$ \citep{sdss_to_ab_magnitude}.}
	\label{tab:mag_cuts}
\end{table}

\subsection{Model Techniques}
\label{sec:machine_learning_method} 

\subsubsection{Feature Selection Process}
\label{sec:feature_selection_process}

It should be noted that we have a total of 32,768 possible combinations for the input features (see the filters and colours described in \S\S\ref{sec:prepare_photometry_from_catalogues}) that could be tested. Due to the computational cost involved to examine all these combinations, we decide to employ an automated feature selection technique known as Sequential Forward Selection (SFS, \citealt{feature_selection}) to determine appropriate filters and colours. This technique is a `greedy' iterative strategy that builds a subset of features via a bottom-up selection approach starting from an empty feature subset. Each iteration evaluates the performance of feature combinations, where SFS selects and stores the feature that best satisfies an objective function\footnote[7]{An objective function is a general term used to describe a function of defined conditions that is minimised or maximised to find the optimal solution for the given objective \citep{objective_function}.} into the empty feature subset. From which, we employ a multi-objective function that checks if the following conditions are satisfied in each iteration of SFS:
    
\begin{enumerate}[leftmargin=0.3cm]
\item The formula below calculates the photometric redshift prediction error: 
\begin{equation}
E_{z} = \frac{|P_{i} - A_{i}|}{(1 + A_{i})} ,
\end{equation}

where $E_{z}$ is the photometric redshift prediction error for each tested cluster, $P_{i}$ is the estimated photometric redshift for each cluster and $A_{i}$ is the `actual'\footnote[8]{This depended on which dataset was used as the photometric redshifts of clusters in the MWAR and WNMR datasets were from the WHL12 cluster catalogue whilst photometric redshifts of clusters in the RNMW dataset were from the redMaPPer cluster catalogue.} photometric redshift for each cluster. Figure SA3 (available online) shows a direct comparison of the photometric redshifts for cross-matched clusters from the WHL12 and redMaPPer cluster catalogues, where both catalogues appear to be in good agreement.

\hspace*{3mm} The median of photometric redshift prediction errors produced during an iteration must be lower than the median of photometric redshift prediction errors from the previous iteration to continue SFS iterations.

\item Filter magnitude-cuts are used to remove galaxies fainter than a specified magnitude threshold for each photometry filter to improve the signal-to-noise of the datasets. This can result in clusters with no galaxies remaining. We determine a percentage of clusters retained by counting the number of clusters that have galaxies remaining, after filter magnitude-cuts are applied, from the initial total in a dataset. From which, we set a threshold for the percentage of clusters retained in the MWAR dataset must be equal or greater than 95 per cent\footnote[9]{A tolerable percentage of data purposely excluded from the dataset should be low, otherwise systematic biases and sample misrepresentation induced by the missing data could be introduced into our analysis \citep{missing_data}.} to continue SFS iterations. 
\end{enumerate}
    
\begin{figure}
\begin{center}
\includegraphics[width=0.75\linewidth, height=0.65\linewidth]{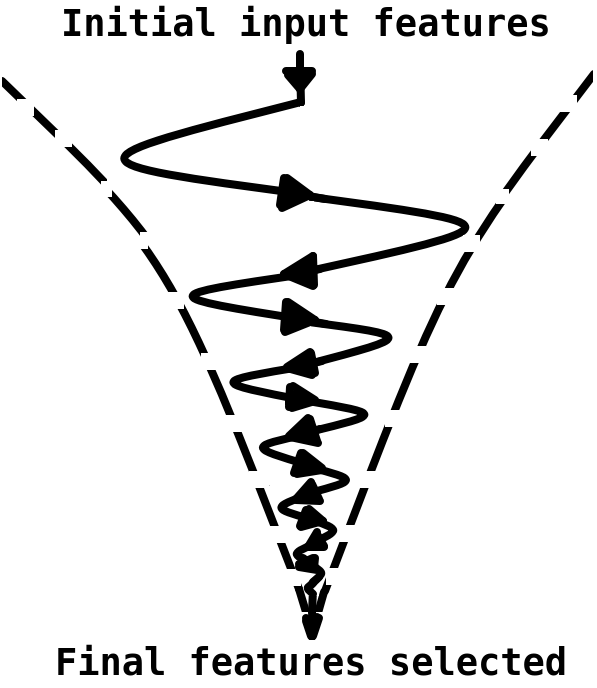}
\caption{A simplified perspective of the SFS strategy. The solid line with black arrows indicate the path taken by SFS to select features and the dashed lines represent the boundaries of feature space. It can be seen that as SFS progresses the feature space shrinks due to the reduced number of possible outcomes, where SFS continues until it converges on a set of features. This diagram was inspired by \protect\cite{SFS_diagram}.}
\label{fig:sequential_forward_selection}
\end{center}
\end{figure}

In Figure \ref{fig:sequential_forward_selection} we observe that the SFS strategy is a computationally efficient approach as it searches through a reduced number of possible combinations, where all selected features are not included for reconsideration in subsequent SFS iterations. The process continues until the objective function is no longer satisfied with the remainder of the input features. We also compare the performance of these features to a control group of features that are not selected with SFS, where the control group features are \textit{g, r, i, g-r, r-i, g-i}. We assume that the control group features would perform well since these filters and colours would likely display `red sequence'-like features over a wide range of redshifts in CMS (\citealt{red_sequence_evolution}; \citealt{redMapper}) accounting for the shifting of the 4000 \AA \ break \citep{4000_angstrom_break_evolution_0}.

\subsubsection{Machine Learning Algorithm}
\label{sec:photoz_algorithm}

\begin{figure*}
	\includegraphics[width=0.85\linewidth, height=0.55\linewidth]{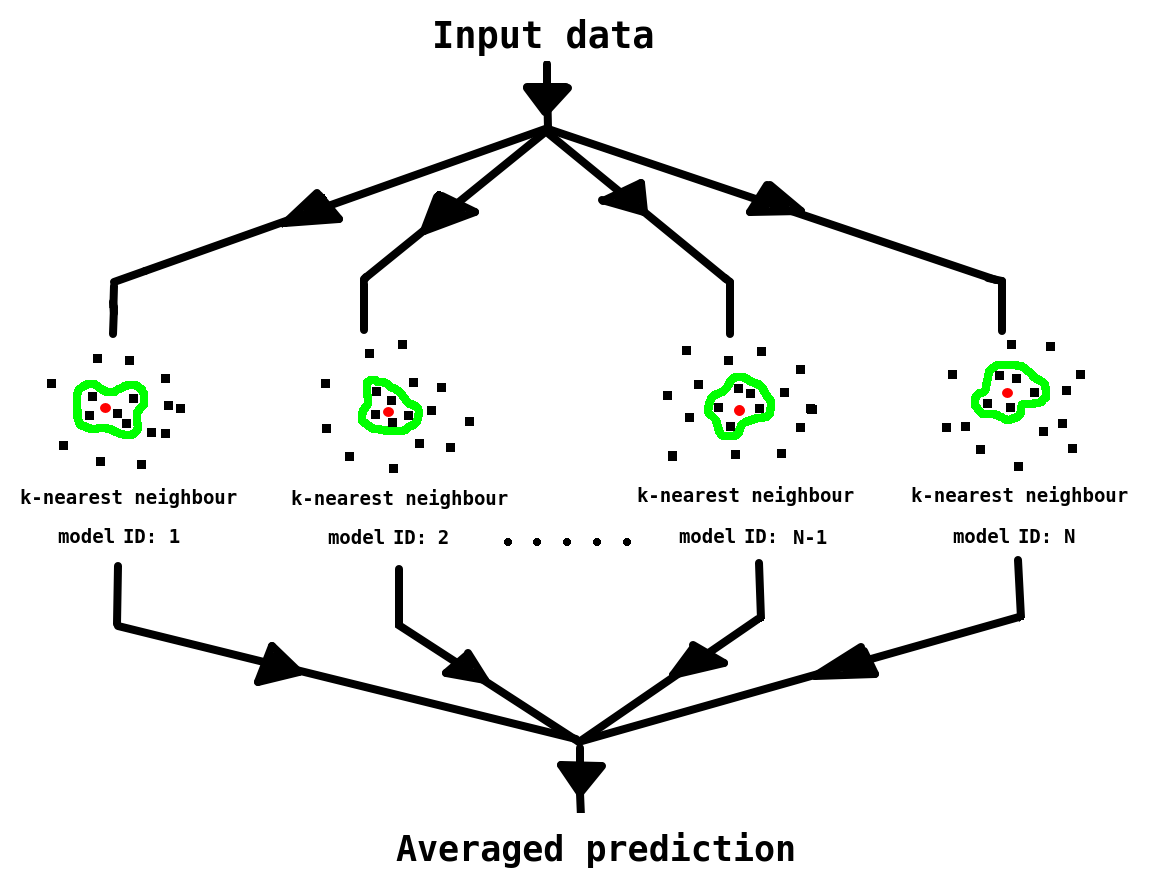}
    \caption{A schematic diagram of the SRKNN algorithm. The solid lines with black arrows indicates the flow of input data to an `N' number of internal KNN models. In this example diagram, we use a red circle in each internal KNN model to represent an input test data point, black squares represent training data points and the green outline show the nearest neighbour training data points from the input test data point. From which, the median of training label values for the corresponding nearest neighbour training data points is used as a prediction for an internal KNN model, where the global model prediction is approximated with the median of predictions across all internal KNN models.}
    \label{fig:sequential_random_knn}
\end{figure*}

We adopt the sequential random k-nearest neighbours (SRKNN, \citealt{srknn}) algorithm as the foundation of our model. The SRKNN algorithm is an ensemble \citep{ensemble_methods} that aggregates multiple k-nearest neighbours (KNN, \citealt{knn_0}; \citealt{knn_1}) models into one global model (see Figure \ref{fig:sequential_random_knn}). The KNN algorithm is classed as a non-parametric learning method \citep{lazy_learning} in the field of machine learning that can be used for non-linear regression tasks. This means that the algorithm has no learnable parameters to train (e.g. weights in a neural network algorithm, \citealt{neural_network}). Predictions for the KNN algorithm are produced by averaging the labelled values of the nearest neighbour training data points to the input data points, where we use the Euclidean distance metric\footnote[10]{It is known that distance comparisons in Euclidean space can become less effective with increasing dimensionality as the distance ratios become more uniform \citep{euclidean_high_dimensions}. This means that other distance metrics such as cosine, Chi-squared, Manhattan and Minkowski \citep{knn_distance_metric} could also be considered.} to compute distances. The main characteristics of the SRKNN algorithm involve bootstrap with replacement (\citealt{bootstrapping_0}; \citealt{bootstrapping_1}) of the training set and random initialisation of input features to train each internal KNN model. These traits can improve the overall accuracy of predictions as a greater variety of features would be considered for each internal KNN model.

The SRKNN algorithm has three main hyper-parameter settings that should be optimised before deployment. These hyper-parameter settings are listed as follows:

\begin{itemize}[leftmargin=0.3cm]
\item The number of internal KNN models (also equivalent to number of bootstrap resamples used). 
\item The number of randomly initialised input features.
\item The number of nearest neighbours.
\end{itemize}

\cite{srknn} suggests that the performance of the SRKNN algorithm depends on the values assigned for each hyper-parameter setting, where the optimal values vary for different datasets. In \S\S\ref{sec:hyper-parameter_analysis} we examine and tune each hyper-parameter setting with the MWAR validation set.

\subsection{Outline Of Model Training}
\label{sec:outline_of_approach} 

Here, we describe the steps used to train and test our model for each search radius. The key points are summarised as follows:

\begin{enumerate}[leftmargin=0.3cm,label*=\arabic*.]

\item Candidate clusters from the WHL12 and redMaPPer cluster catalogues were split into training, validation and test sets. The MWAR dataset was designated as the training/validation set (80:20 per cent split ratio), whilst the RNMW/WNMR datasets were used as test sets. Photometric measurements of observed galaxies in the clusters was obtained from the SDSS-III DR9 photometric catalogue and full-sky dust reddening maps (\citealt{dust_extinction_0}; \citealt{dust_extinction_1}) were also used to account for galactic extinction. \newline

\item All the filters and colours described in \S\S\ref{sec:prepare_photometry_from_catalogues} are assigned as input features to a single KNN algorithm for feature selection and filter magnitude-cut analysis. If a filter was used as part of an input feature, then the corresponding filter magnitude-cut was applied to exclude galaxies that had poor photometric measurements in that filter. The mean and standard deviation were also calculated for each feature in the MWAR training set to perform feature scaling\footnote[11]{All photometric measurements of features are standardised with zero-mean centering and unit variance, which is necessary for the comparison of Euclidean distance measurements \citep{feature_scaling}.}. From which, all input datasets to our model will require feature scaling with the same mean and standard deviation values determined for the MWAR training set. \newline

\item Thirty repetitions of ten-fold cross validation \citep{k_fold_cv} were computed with SFS for a individual KNN algorithm, where a single nearest neighbour was used\footnote[12]{A single nearest neighbour minimises algorithmic biases which in turn maximises the variance of predictions \citep{1_nearest_neighbour}.}. This process was important for multiple reasons. Firstly, to analyse the stability of the KNN algorithm from minor changes to the training set. Secondly, to examine the relative frequency of features selected by SFS. Thirdly, to evaluate how filter magnitude-cuts affect the accuracy of photometric redshift predictions. Lastly, to provide a basis for comparing an individual algorithm with an ensemble algorithm. \newline

\item The optimal filter magnitude-cuts determined for a single KNN algorithm were utilised for the SRKNN algorithm via transfer learning \citep{transfer_learning}. From which, the training data for the internal KNN models of the SRKNN algorithm were built with bootstrap resamples, where bootstrap with replacement of the MWAR training set was used. Any clusters that were not used for bootstrapping of an internal KNN model were instead used for feature selection training of that internal KNN model with SFS. This ensured that all available training data was utilised. \newline

\item The hyper-parameter settings of the SRKNN algorithm were tuned via a grid search strategy \citep{grid_and_random_search} using hold-out validation \citep{hold-out_validation} of the MWAR validation set. This also examined how each of the hyper-parameter settings affected the model performance and generalisation. \newline

\item Evaluation of the tuned model performance was obtained with the WNMR/RNMW test sets, which were all unseen clusters. Uncertainties for the photometric redshift estimate of each cluster were approximated with empirical bootstrap confidence intervals. Additionally, the tuned model was run on clusters with low richness\footnote[13]{We define a cluster with low richness as a cluster that has a richness of twenty or fewer member galaxies.} and clusters at high redshift\footnote[14]{We define a cluster at high redshift as a cluster that has a photometric redshift equal or greater than 0.6, which is the upper limit of our training set.} to assess the response of the tuned model on clusters with unseen properties.

\end{enumerate}

\section{Results}
\label{sec:results}

\subsection{Feature Selection and Filter Magnitude-Cut Analysis}
\label{sec:feature_selection_analysis}

\begin{figure*}
	\includegraphics[width=0.32\textwidth]{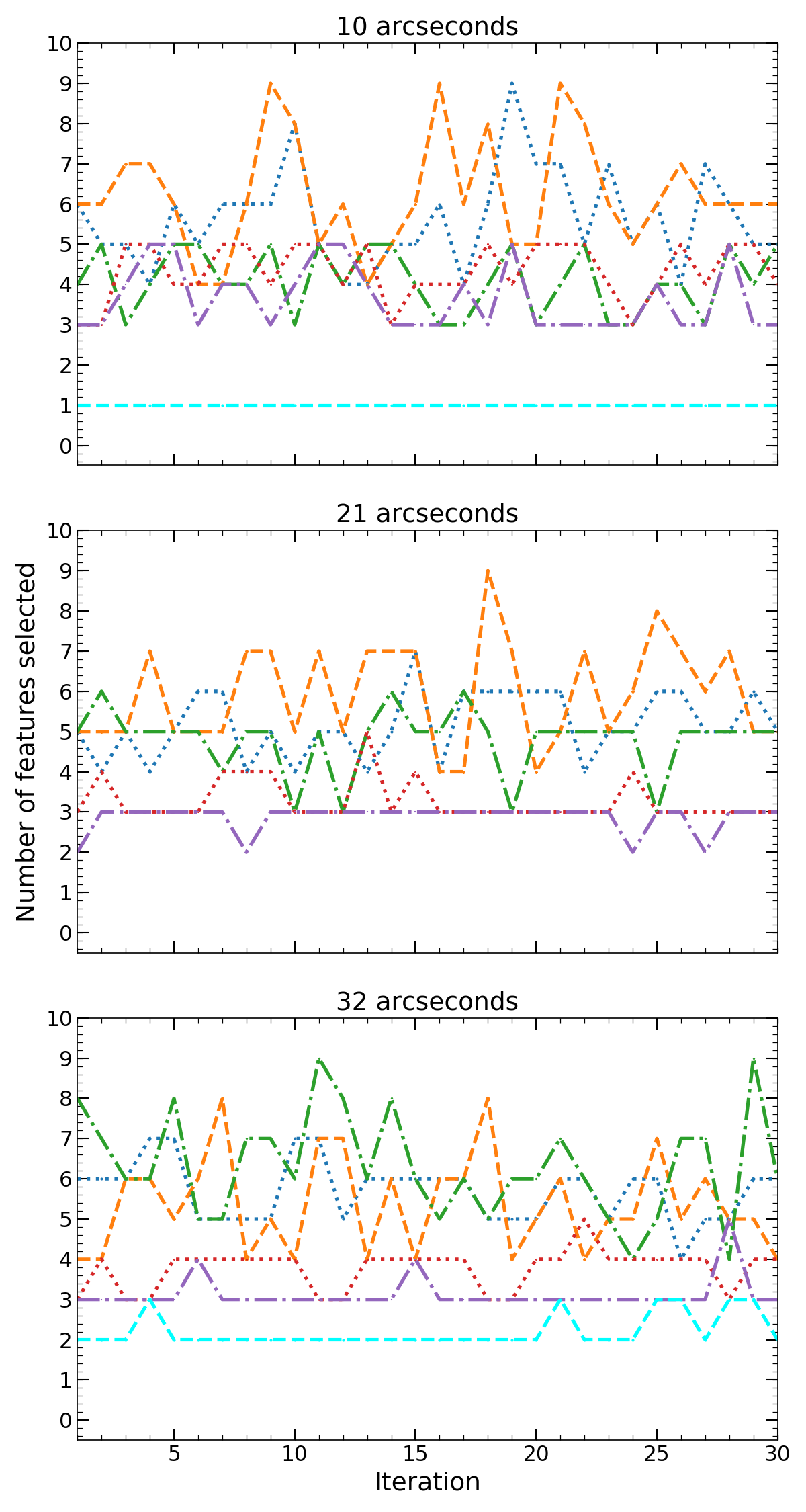}
	\includegraphics[width=0.32\textwidth]{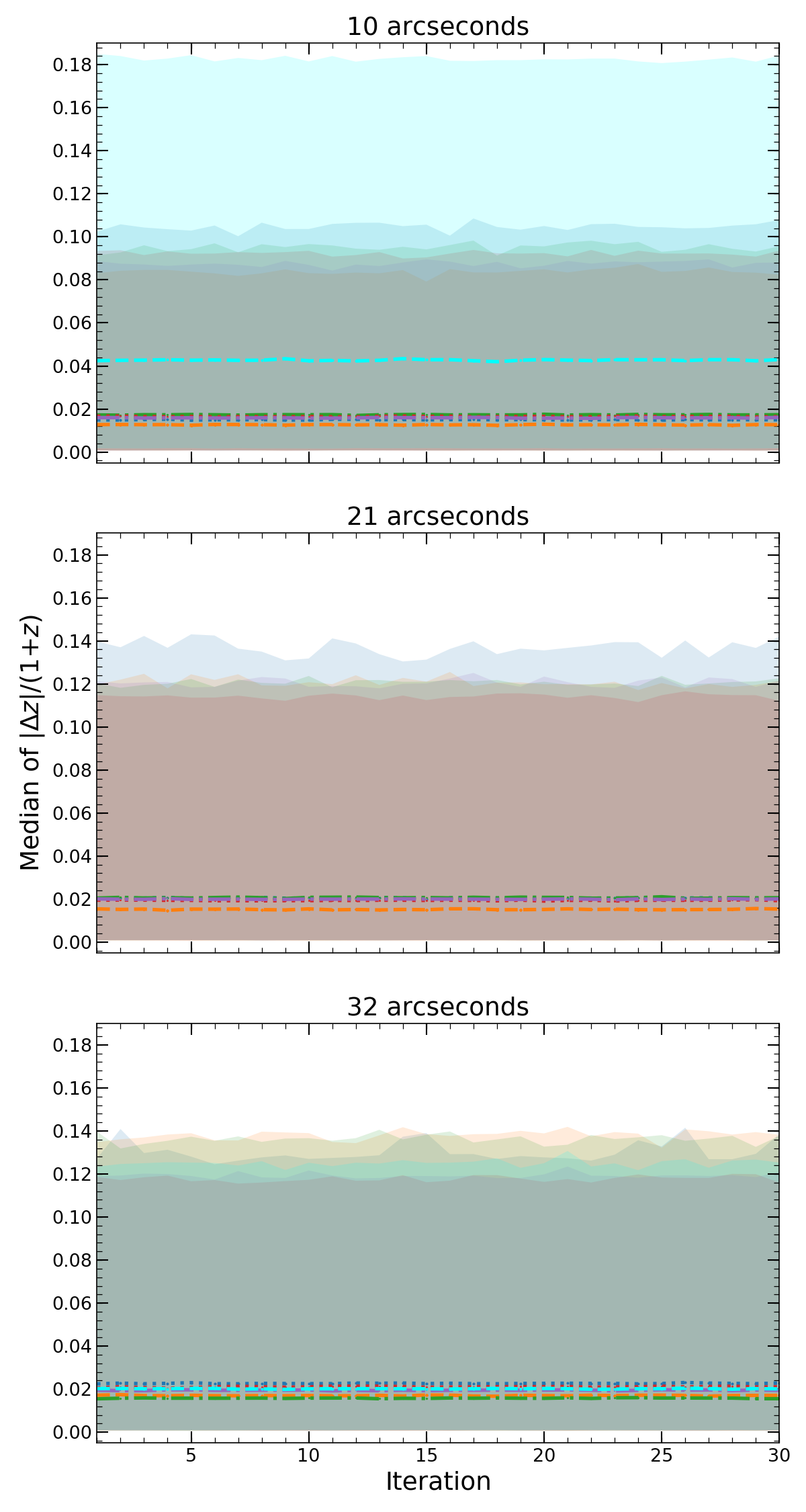}
	\includegraphics[width=0.32\textwidth]{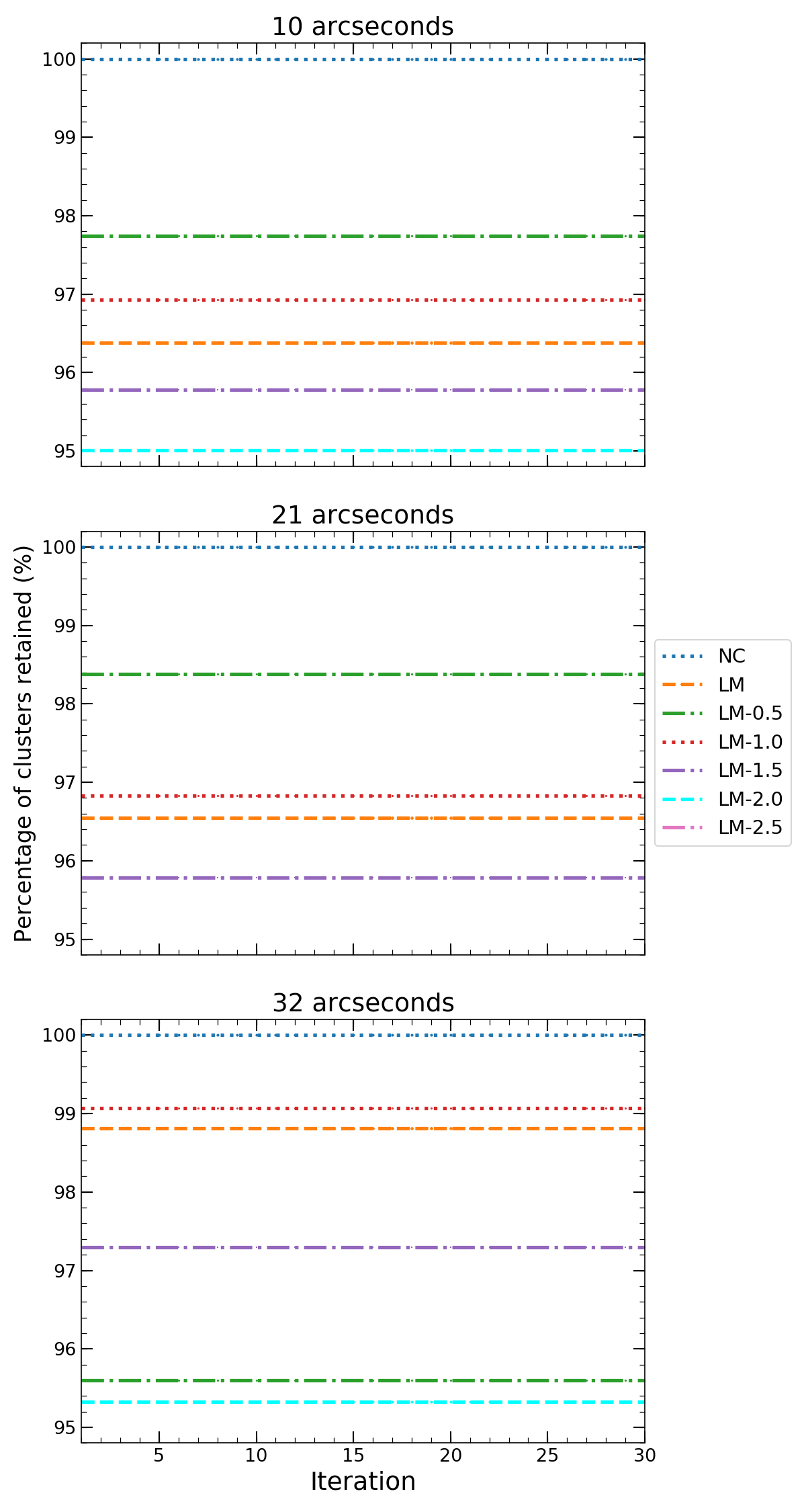}
    \caption{Plots displaying the results from applying filter magnitude-cuts to the MWAR training set using a single KNN algorithm with SFS selected features for each search radii (10 arcseconds on the top row, 21 arcseconds on the middle row and 32 arcseconds on the bottom row). `NC' represents a dataset with no filter magnitude-cuts applied and `LM' represents the MWAR dataset with SFS selected features where filter magnitude-cuts are applied to the limiting magnitude of SDSS. In addition, `LM' is the faintest filter magnitude-cut whilst `LM-2.5' is the brightest filter magnitude-cut. Left column: Number of features selected for the best performing feature subset in ten-fold cross validation across thirty repeats. Middle column: Median of photometric redshift prediction errors ($|\Delta z|/(1+z)$) across all tested clusters for the best performing feature subset in ten-fold cross validation across thirty repeats, where the shaded regions represent 95 per cent confidence intervals. Right column: Percentage of test clusters retained after filter magnitude-cuts are applied with the best performing feature subset in ten-fold cross validation across thirty repeats. It should also be noted that if the percentage of clusters retained, after filter magnitude-cuts are applied, do not satisfy the 95 per cent cluster retainment threshold we would not display the corresponding results in the other columns.}
    \label{fig:feature_frequency_and_predictions}
\end{figure*}

Following the procedure described in \S\ref{sec:outline_of_approach}, we first examine the stability of photometric redshift predictions for a single KNN algorithm. As seen in Figure \ref{fig:feature_frequency_and_predictions}, we observe that for brighter filter magnitude-cuts the number of selected features by SFS are more contrast, such that the resultant feature subsets for fainter filter magnitude-cuts are more strongly influenced by the observations in the MWAR training set itself. However, as seen from the corresponding photometric redshift prediction errors, we find that this did not significantly alter the stability of predictions. We also compare the performance of SFS selected features with the control group features (see \S\S\S\ref{sec:feature_selection_process}), which had not been SFS selected. We repeat the same procedure used to analyse the SFS selected features for the control group features as well. From which, in Figure S1 (available online) we find that the control group features tend to have larger photometric redshift prediction errors in comparison to the SFS selected features for each search radius.

By repeatedly applying ten-fold cross validation to the MWAR training set we could also examine the relative frequency of features selected by SFS. This was done by calculating the relative frequency of features observed in the best performing feature subsets across all thirty repeats. As seen from Table \ref{tab:feature_frequency}, we find that some of the features are frequently selected whilst other features are rarely chosen, such that certain features are more likely to be picked by SFS if they are present in the input features.

\begin{table*}
    \centering
	\begin{tabular}{|c|c|c|c|}
		\hline
	    Search Radius & Optimal Filter Magnitude-Cut & SFS Selected Features & Relative Frequency Of SFS Selected Features \\ \newline
        [arcseconds] & [mag] & & (per cent) \\
		\hline
		10 & LM & \textit{r-i, g-z, r-z, g, g-i, z, r, i-z, g-r, i} & 100, 100, 90, 83, 67, 53, 47, 40, 30, 13 \\
		21 & LM & \textit{z, r-i, g-i, g-z, r, g, g-r, i, r-z} & 87, 80, 80, 70, 63, 60, 60, 47, 47 \\
	    32 & LM-0.5 & \textit{g-z, r-i, g-i, g-r, g, i-z, z, r, i, r-z} & 93, 83, 83, 77, 70, 60, 50, 47, 43, 27 \\
		\hline
	\end{tabular}
	\caption{A table displaying the relative frequency of features selected by SFS across thirty repeats of ten-fold cross validation on the MWAR training set with a single KNN algorithm at the optimal filter magnitude-cut for each search radius. The selected features are listed in the same order as the corresponding relative frequency. It can be seen that the \textit{z} filter, rather than a colour, has the highest relative frequency amongst the features at the 21 arcseconds search radius for a single KNN algorithm but the relative frequency diminishes when the \textit{z} filter is instead used in an ensemble (see \S\S\ref{sec:hyper-parameter_analysis}).}
	\label{tab:feature_frequency}
\end{table*}

Next, we determine the optimal filter magnitude-cut for each search radius by identifying filter magnitude-cut values that returned the lowest photometric redshift prediction error and retained at least 95 per cent of clusters. In Figures \ref{fig:feature_frequency_and_predictions} and S1 (available online), we find that the LM filter magnitude-cut is the optimal filter magnitude-cut for the 10 and 21 arcseconds search radii whilst the LM-0.5 filter magnitude-cut is the optimal filter magnitude-cut for the 32 arcseconds search radius. We also compare whether applying filter magnitude-cuts improves the predictive performance of the model. In Figures \ref{fig:feature_frequency_and_predictions} and S1 (available online) we find that a dataset, NC, with no filter magnitude-cuts applied to it, is not the optimal filter magnitude-cut for any search radius whilst datasets with filter magnitude-cuts applied often had lower photometric redshift prediction errors.

\begin{figure}
\begin{center}
	\includegraphics[width=0.95\linewidth, height=0.85\linewidth]{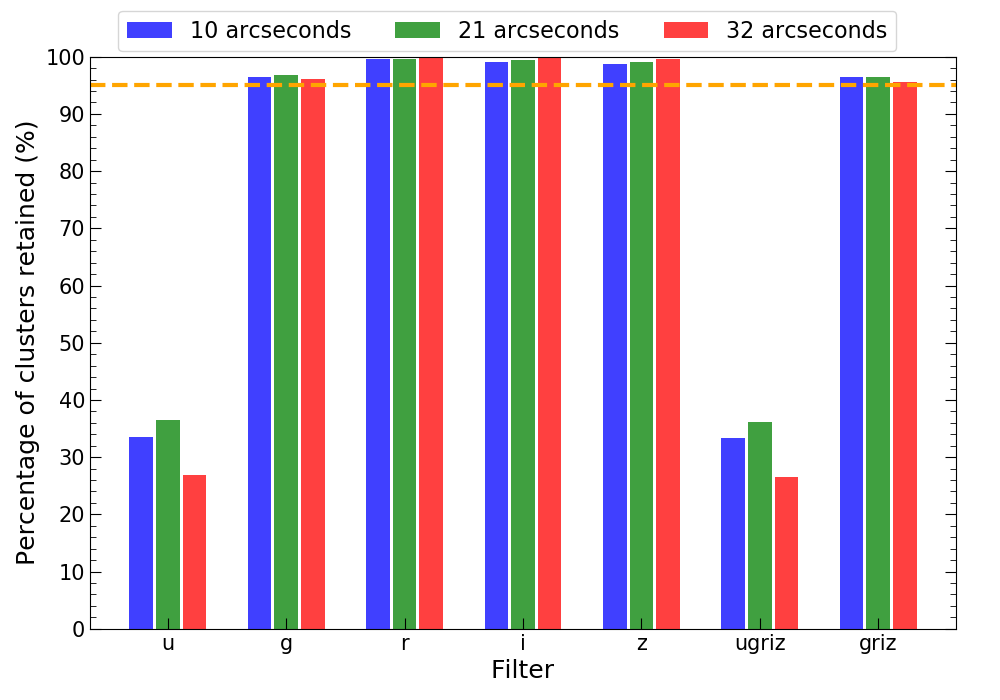}
    \caption{Percentage of clusters retained in the MWAR training set after applying the optimal filter magnitude-cuts for each search radius to the \textit{u}, \textit{g}, \textit{r}, \textit{i}, \textit{z}, \textit{ugriz} and \textit{griz} filters. The orange dashed line highlights the 95 per cent cluster retainment threshold.}
    \label{fig:filter_mag_cut}
\end{center}
\end{figure}

We also assess how magnitude-cuts of the filters themselves affect the percentage of clusters retained in the MWAR training set, where the optimal filter magnitude-cut for each search radius was applied. From Figure \ref{fig:filter_mag_cut} we find that all filters, except for the \textit{u} filter, satisfied the 95 per cent cluster retainment threshold at each search radius. In addition, we observe in Table \ref{tab:feature_frequency} that the \textit{u} filter did not appear in any final feature subset. From which, we decide that all input features which did not involve the \textit{u} filter would be used as the new input features for the SRKNN algorithm to reduce the computational cost of evaluating redundant features during feature selection training. One would expect the \textit{u} filter to be a poor predictor of redshift beyond very low redshift as it will probe further into the UV with increased redshift.

\subsection{Hyper-Parameter Tuning Analysis Of The SRKNN Algorithm}
\label{sec:hyper-parameter_analysis}

We combine the optimal filter magnitude-cuts learned in \S\S\ref{sec:feature_selection_analysis} with a grid search strategy to fine-tune the SRKNN algorithm, which is known as inductive transfer learning (\citealt{inductive_transfer_learning_0}; \citealt{inductive_transfer_learning_1}). We assume that the knowledge learned for the KNN algorithm is appropriate for the SRKNN algorithm, since the SRKNN algorithm is an extension of the KNN algorithm. From which, we ran the grid search on all combinations of hyper-parameter settings with a specified range of values to evaluate how each hyper-parameter setting affects model generalisation and predictive performance. The following hyper-parameter setting values are used in the grid search: 

\begin{itemize}[leftmargin=0.3cm]
\item The number of internal KNN models - 1, 10, 20, 30, 40, 50, 60, 70, 80, 90, 100, 150, 200, 250, 300, 350, 400, 450, 500, 550, 600, 650, 700, 750, 800, 850, 900, 950, 1000.
\item The number of initialised random features - 1, 2, 3, 4, 5, 6, 7, 8, 9, 10.
\item The number of nearest neighbours - 1, 2, 3, 4, 5, 6, 7, 8, 9, 10, 11, 12, 13, 14, 15, 16, 17, 18, 19, 20, 21, 22, 23, 24, 25.
\end{itemize}

We utilise validation curves \citep{validation_curves} to analyse the response from different hyper-parameter setting combinations of the SRKNN algorithm. This involves fixing each hyper-parameter setting as a constant with respect to the other hyper-parameter settings to compute the median of photometric redshift prediction errors across all tested clusters with that fixed hyper-parameter setting. We focus on minimising the photometric redshift prediction error on the MWAR validation set rather than the MWAR training set. Since the MWAR training set had already been seen by the model, the results from the MWAR training set would be biased whilst the MWAR validation set remains unseen by the model. However, running the model on both the MWAR training and validation sets is still beneficial to check the generalisation of the hyper-parameter settings, as the model can overfit and underfit when applied on its own training data.

\begin{figure}
	\includegraphics[width=0.49\textwidth]{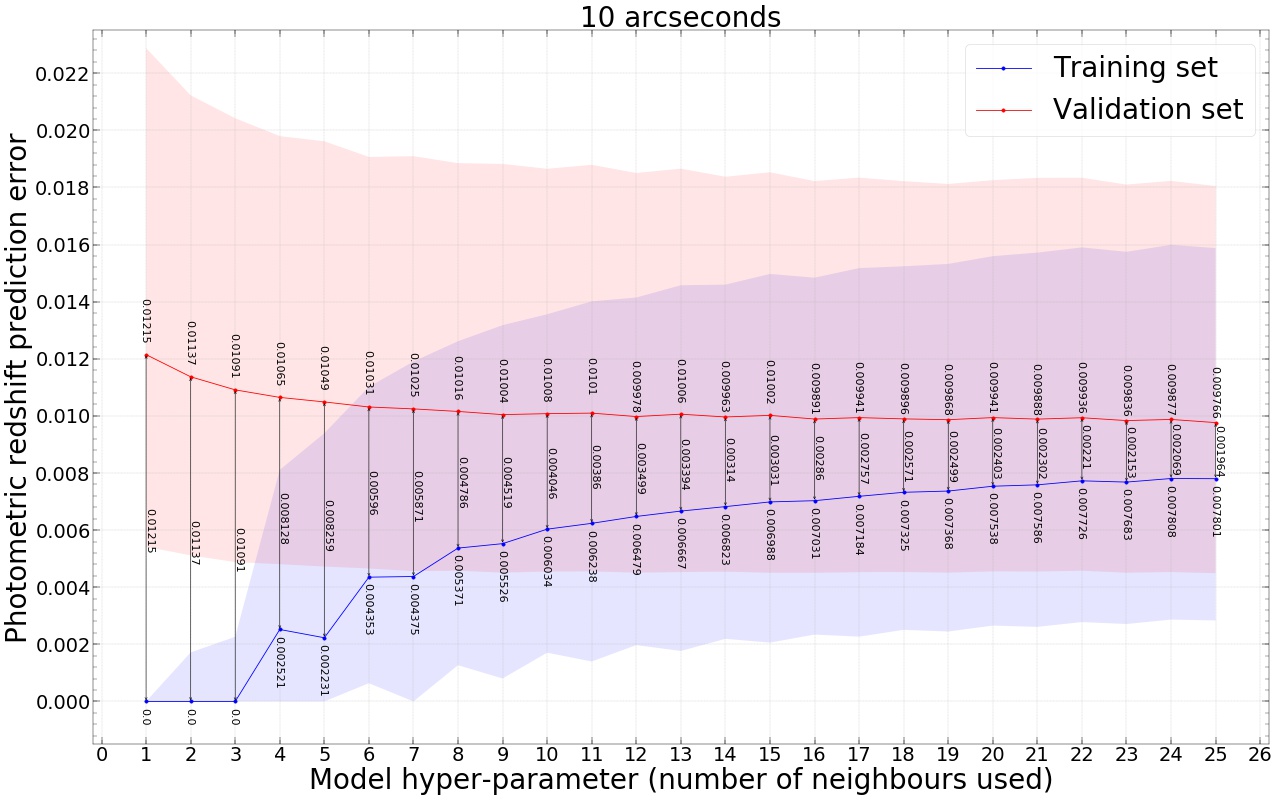}
	\includegraphics[width=0.49\textwidth]{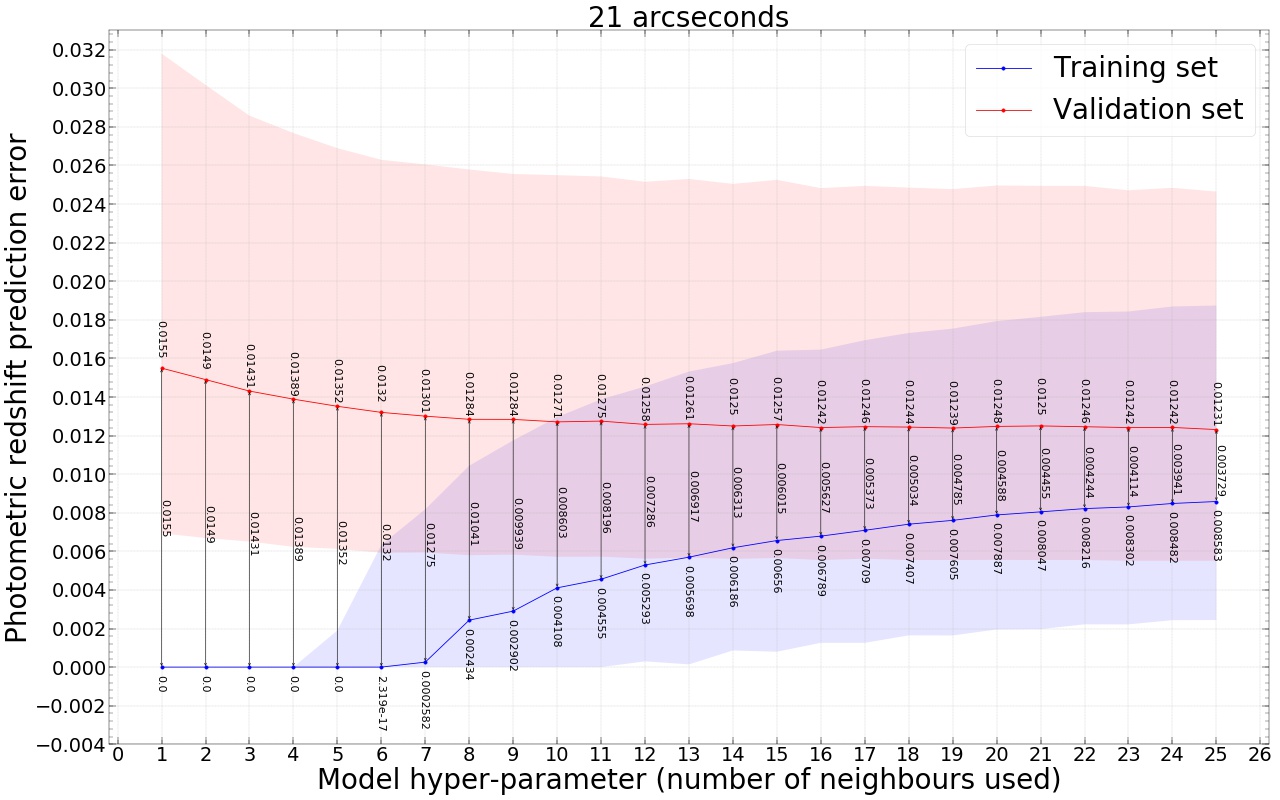}
	\includegraphics[width=0.49\textwidth]{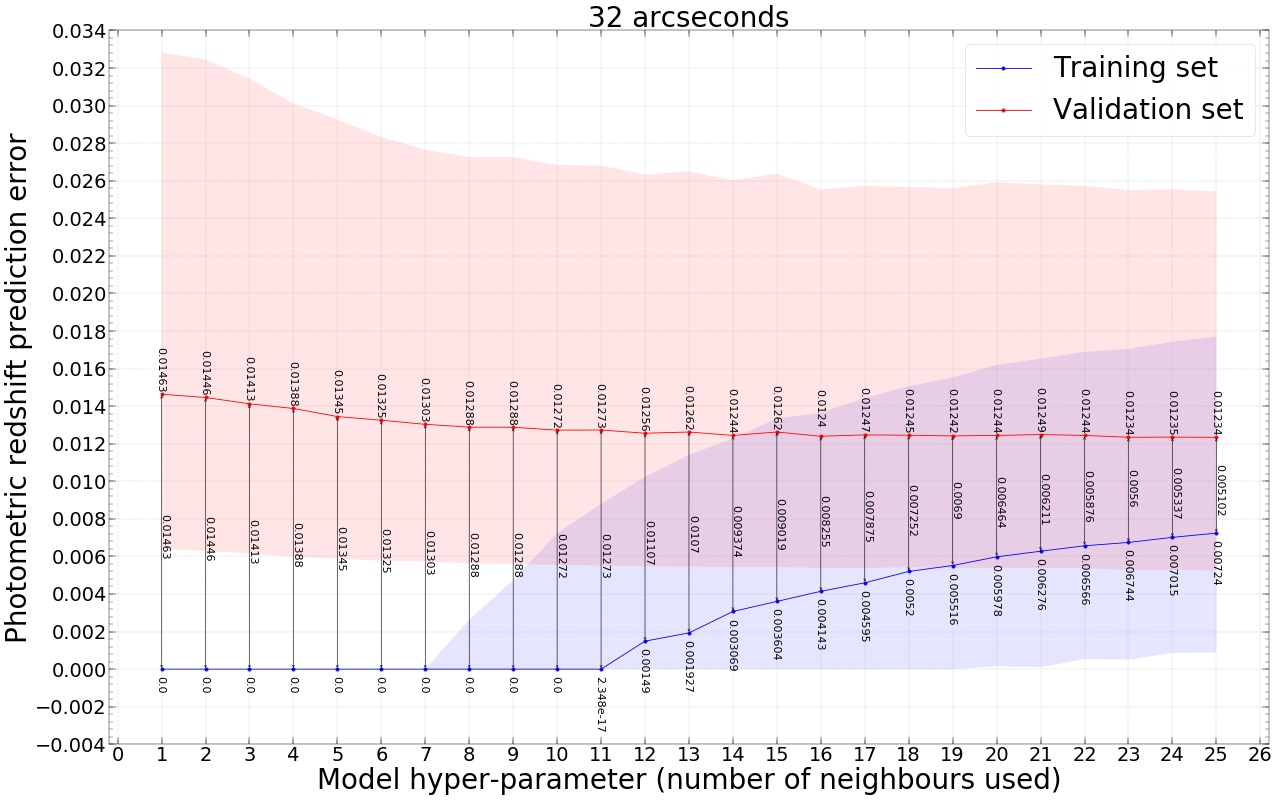}
    \caption{Validation curves from tuning the number of nearest neighbours hyper-parameter setting, where the photometric redshift prediction errors of the MWAR training (blue) and validation (red) sets are shown for each search radii  (10 arcseconds on the top row, 21 arcseconds in the middle row and 32 arcseconds on the bottom row). The individual points display the median of photometric redshift prediction errors across all tested clusters and the shaded regions represent the 25th and 75th percentiles of the photometric redshift prediction errors for a fixed number of nearest neighbours with respect to the other hyper-parameter settings of the SRKNN algorithm. We also label the difference between the individual points of the training and validation errors.}
    \label{fig:validation_curve_k}
\end{figure}

Firstly, we evaluate the model performance based on the number of nearest neighbours for each search radius. In Figure \ref{fig:validation_curve_k}, we find that for a small number of nearest neighbours the model has high predictive variance as we observe a large difference between the training and validation errors. Although, we notice that the overall photometric redshift prediction error decreases as the number of nearest neighbours increases for the MWAR validation set, whereas the overall photometric redshift prediction error increases as the number of nearest neighbours increases for the MWAR training set. It can be seen that the number of nearest neighbours is a very important hyper-parameter setting to tune since the model performance varies a lot depending on the value used. From which, we determine the optimal values for the number of nearest neighbours of each search radius to be 19 for 10 arcseconds, 19 for 21 arcseconds and 16 for 32 arcseconds. It should be noted that the number of nearest neighbours value with the lowest photometric redshift prediction error was actually 25 for each search radius. We purposely avoid selecting this value since the number of nearest neighbours value has a large impact on the model performance, such that selecting the hyper-parameter value with the lowest photometric prediction error could likely overfit the model on the MWAR validation set itself. Instead, we prefer to choose more conservative values for the optimal number of nearest neighbours to balance model generalisation and performance.

\begin{figure}
	\includegraphics[width=0.49\textwidth]{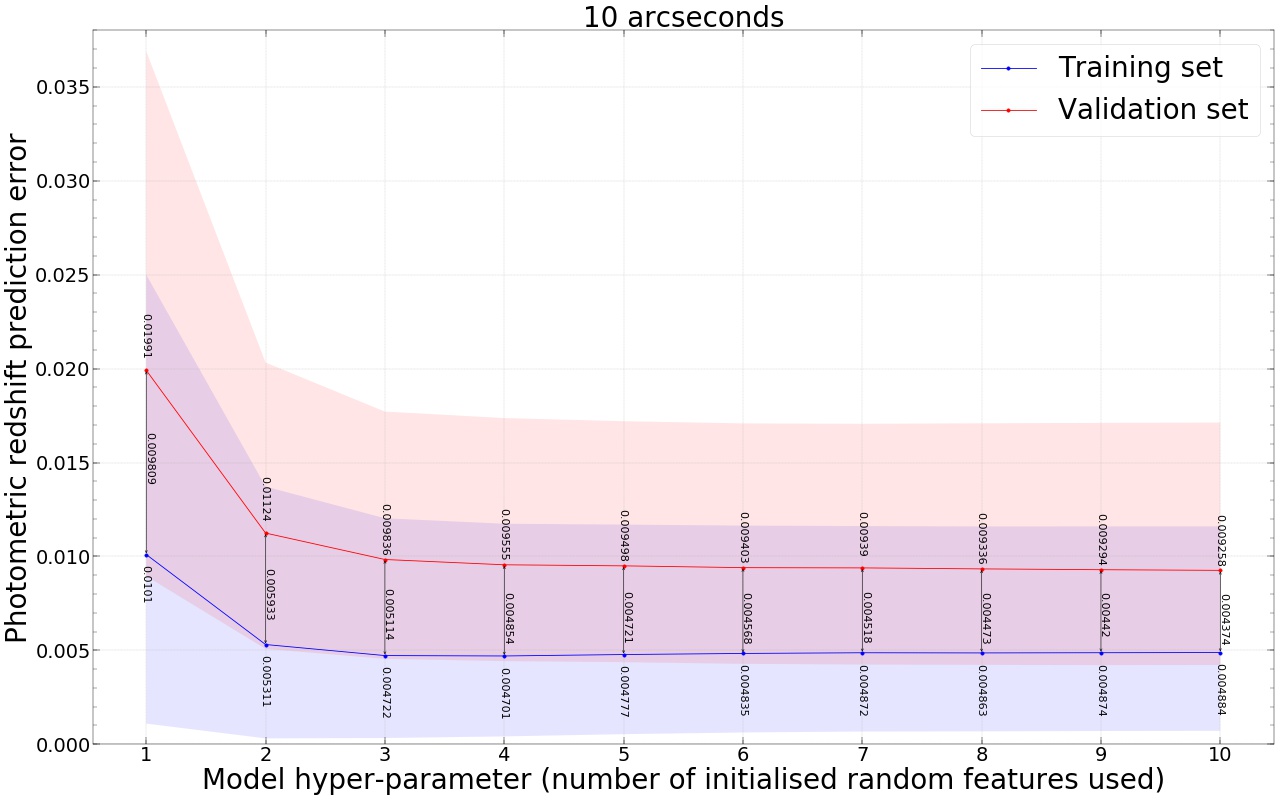}
	\includegraphics[width=0.49\textwidth]{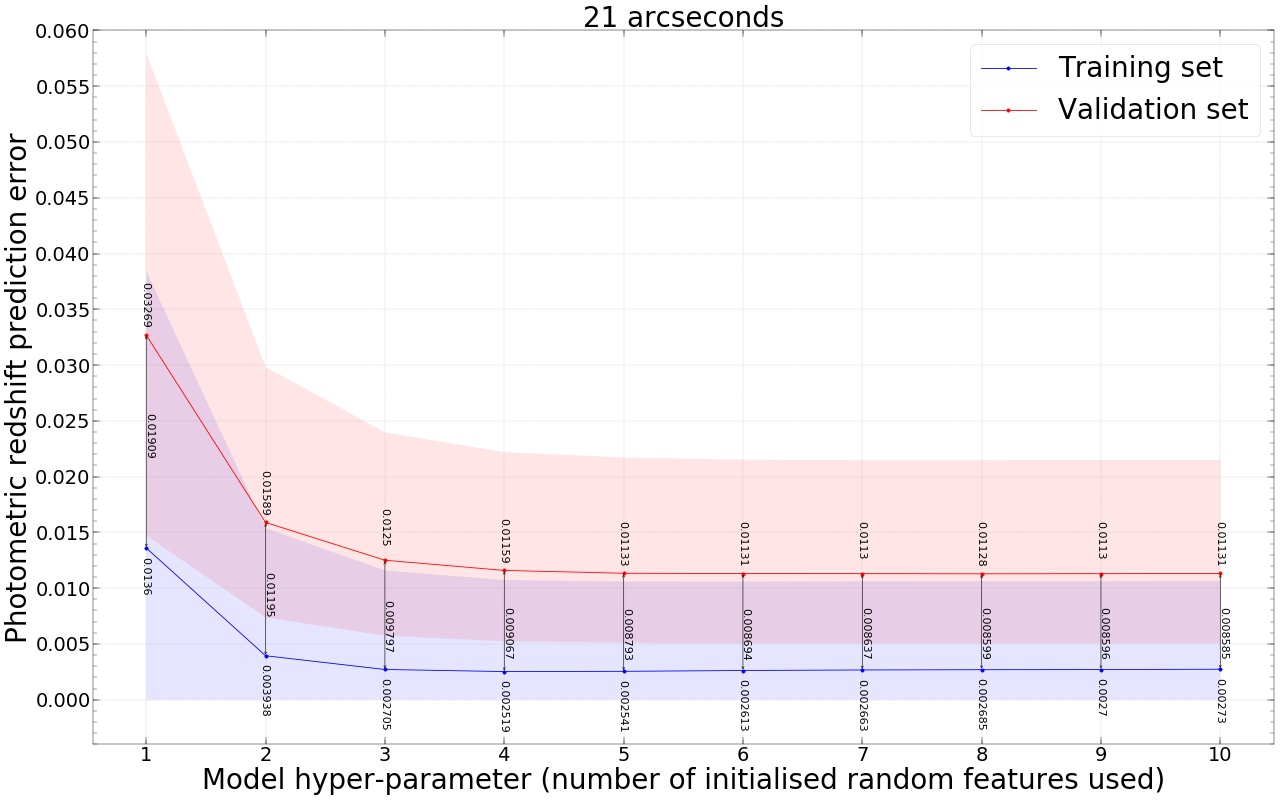}
	\includegraphics[width=0.49\textwidth]{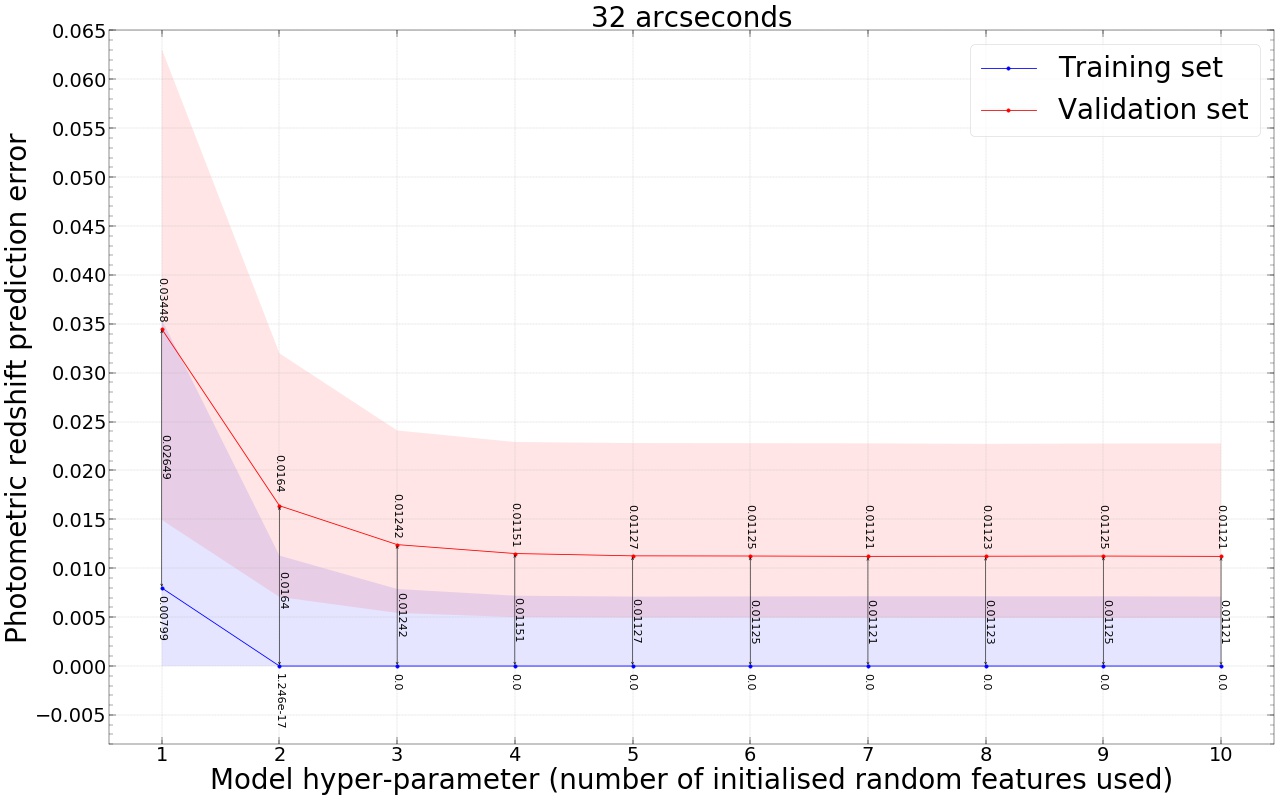}
    \caption{This figure is equivalent to Figure \ref{fig:validation_curve_k} except we tune the number of initialised random features hyper-parameter setting.}
    \label{fig:validation_curve_features}
\end{figure}

Secondly, we examine the model performance based on the number of initialised random features for each search radius. In Figure \ref{fig:validation_curve_features}, we find that for both the MWAR training and validation sets, the change in the photometric redshift prediction errors quickly decreases for a small number of initialised random features but then slowly decreases when a medium to large number of initialised random features was used. From which, we observe that the overall redshift prediction error decreases as the number of initialised random features increases. This implies that the number of initialised random features is also an important hyper-parameter setting to tune, since the model performance on the MWAR training and validation sets is somewhat reliant on the value selected. We determine the optimal values for the number of initialised random features of each search radius to be 9 for 10 arcseconds, 8 for 21 arcseconds and 7 for 32 arcseconds. Although, it can be seen that having no initialised random features (using all features for the input features) at times had lower photometric redshift prediction errors. However, this could also worsen model generalisation since strongly correlated features would not be restricted during SFS. Therefore, we again decide to select more conservative values for the optimal number of initialised random features.

\begin{figure}
	\includegraphics[width=0.49\textwidth]{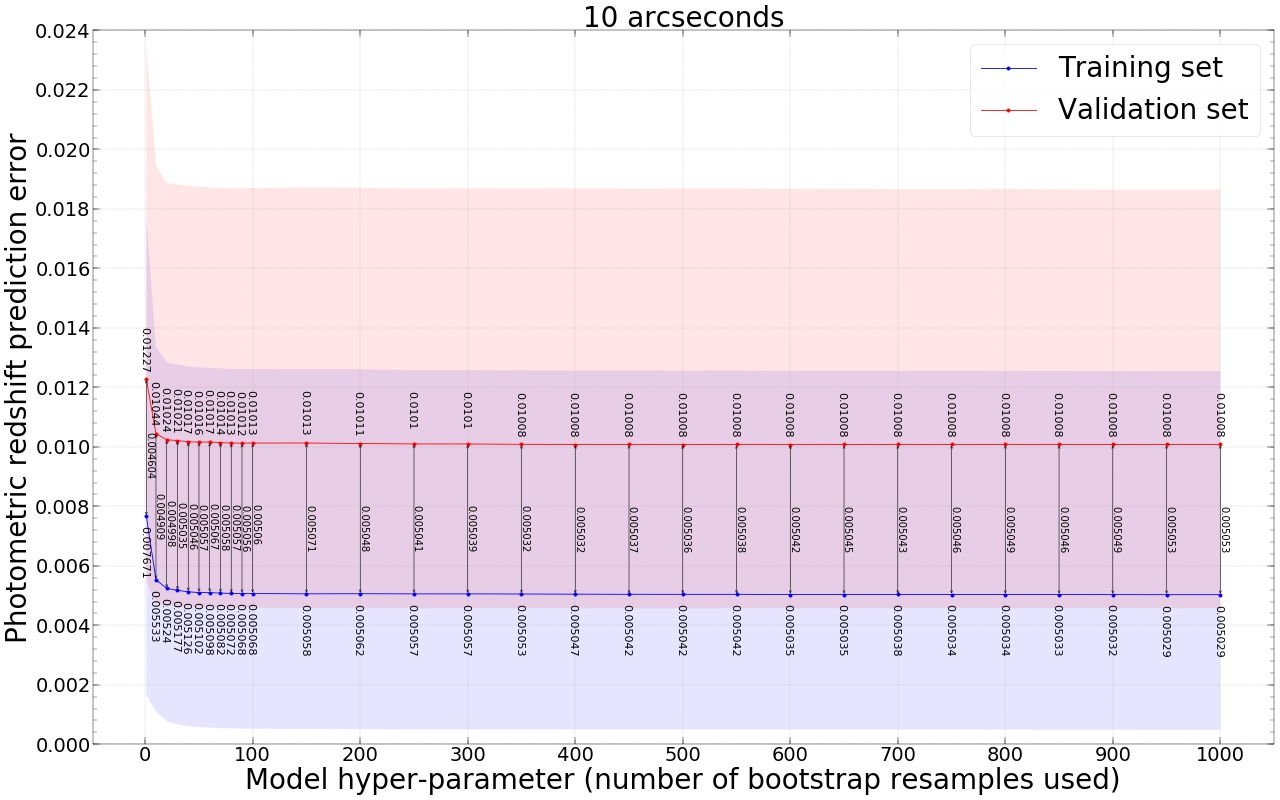}
	\includegraphics[width=0.49\textwidth]{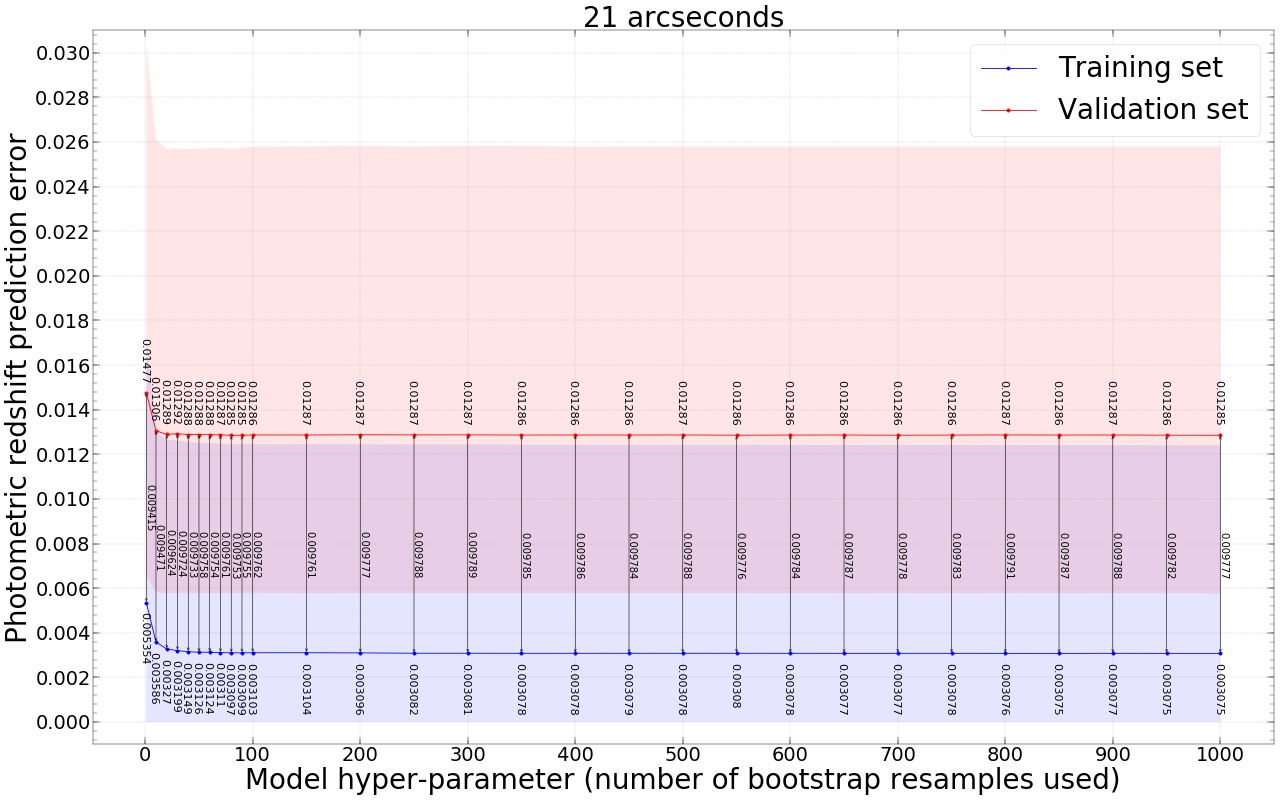}
	\includegraphics[width=0.49\textwidth]{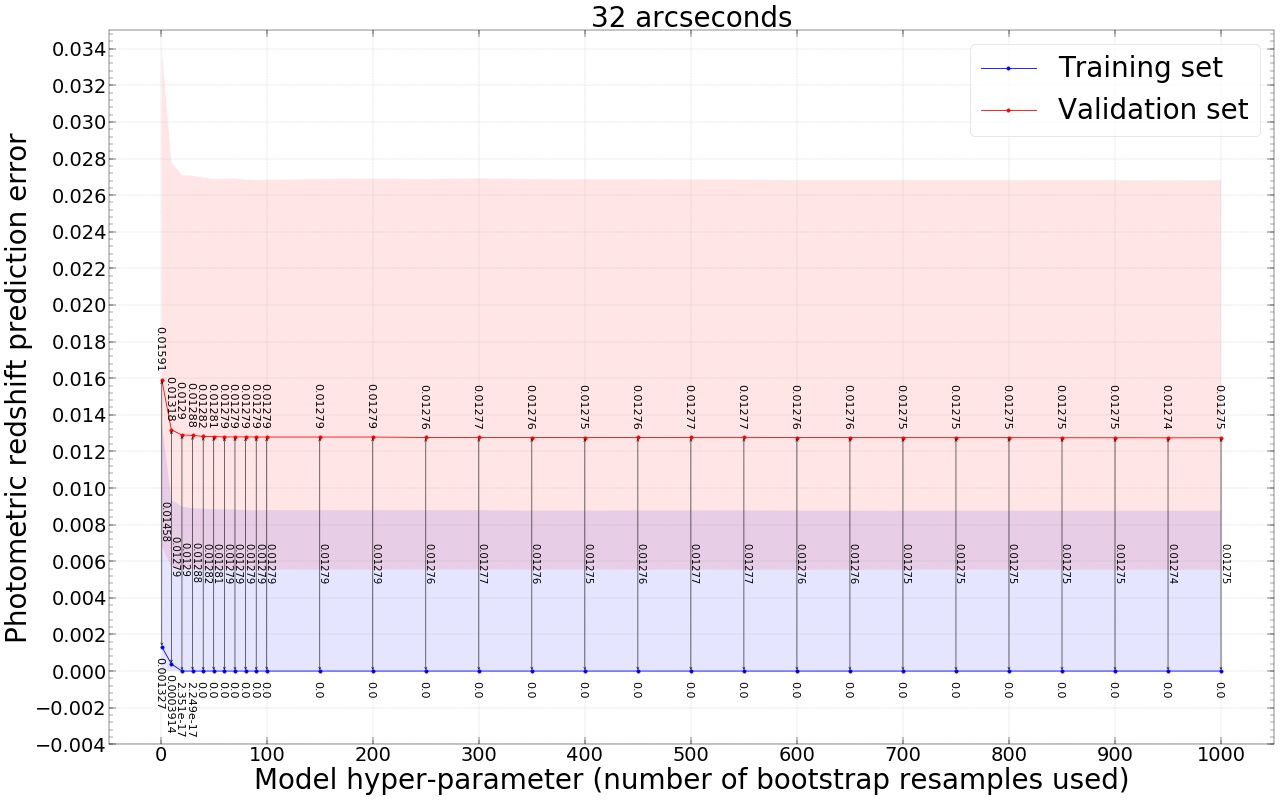}
    \caption{This figure is equivalent to Figure \ref{fig:validation_curve_k} except we tune the number of bootstrap resamples hyper-parameter setting.}
    \label{fig:validation_curve_bootstraps}
\end{figure}

\begin{figure*}
	\includegraphics[width=\linewidth]{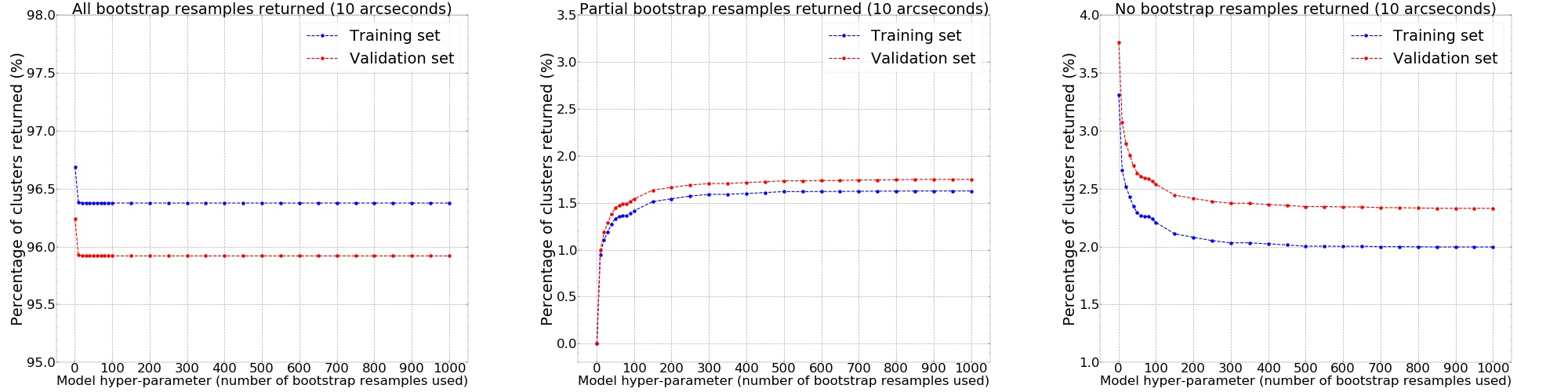}
	\includegraphics[width=\linewidth]{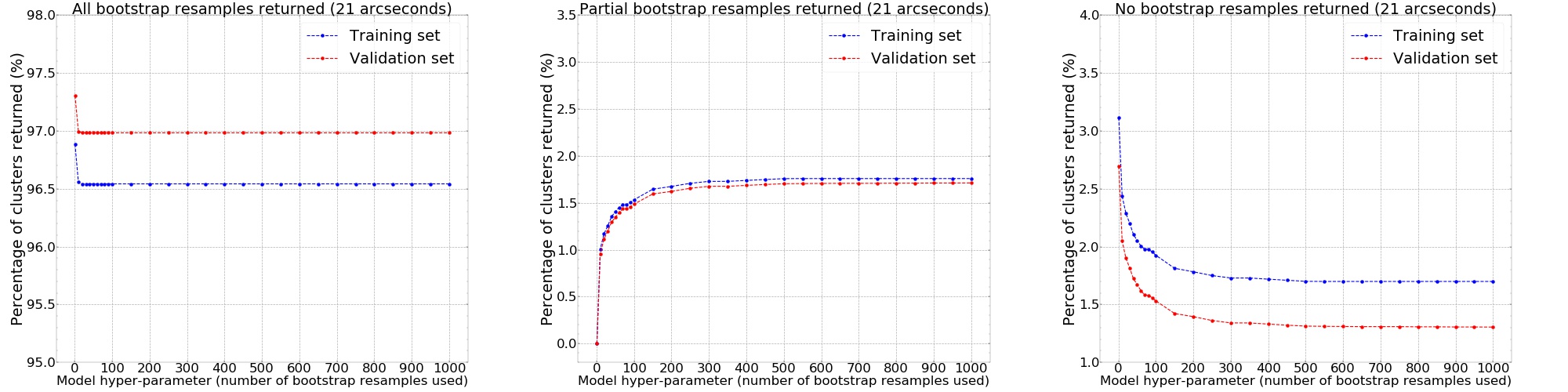}
	\includegraphics[width=\linewidth]{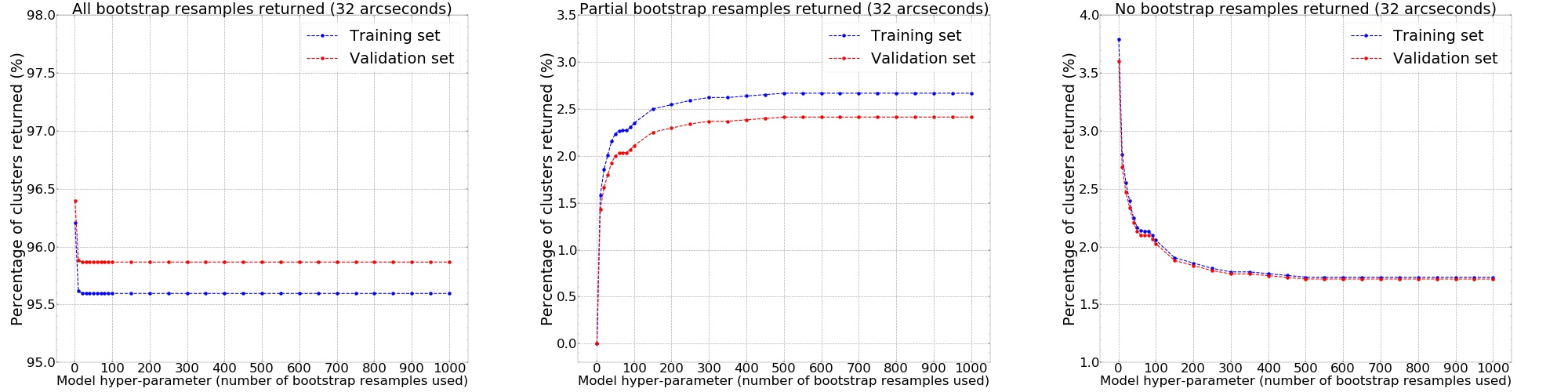}
    \caption{Validation curves from tuning the number of bootstrap resamples hyper-parameter setting, where the percentage of clusters returned with full, partial and no bootstrap resamples are from the MWAR training (blue) and validation (red) sets at each search radii (10 arcseconds on the top row, 21 arcseconds in the middle row and 32 arcseconds on the bottom row). The individual points display the percentage of clusters returned across a fixed number of bootstrap resamples with respect to the other hyper-parameter settings of the SRKNN algorithm.}
    \label{fig:validation_curve_bootstraps_percentages}
\end{figure*}

\begin{figure}
	\includegraphics[width=0.49\textwidth]{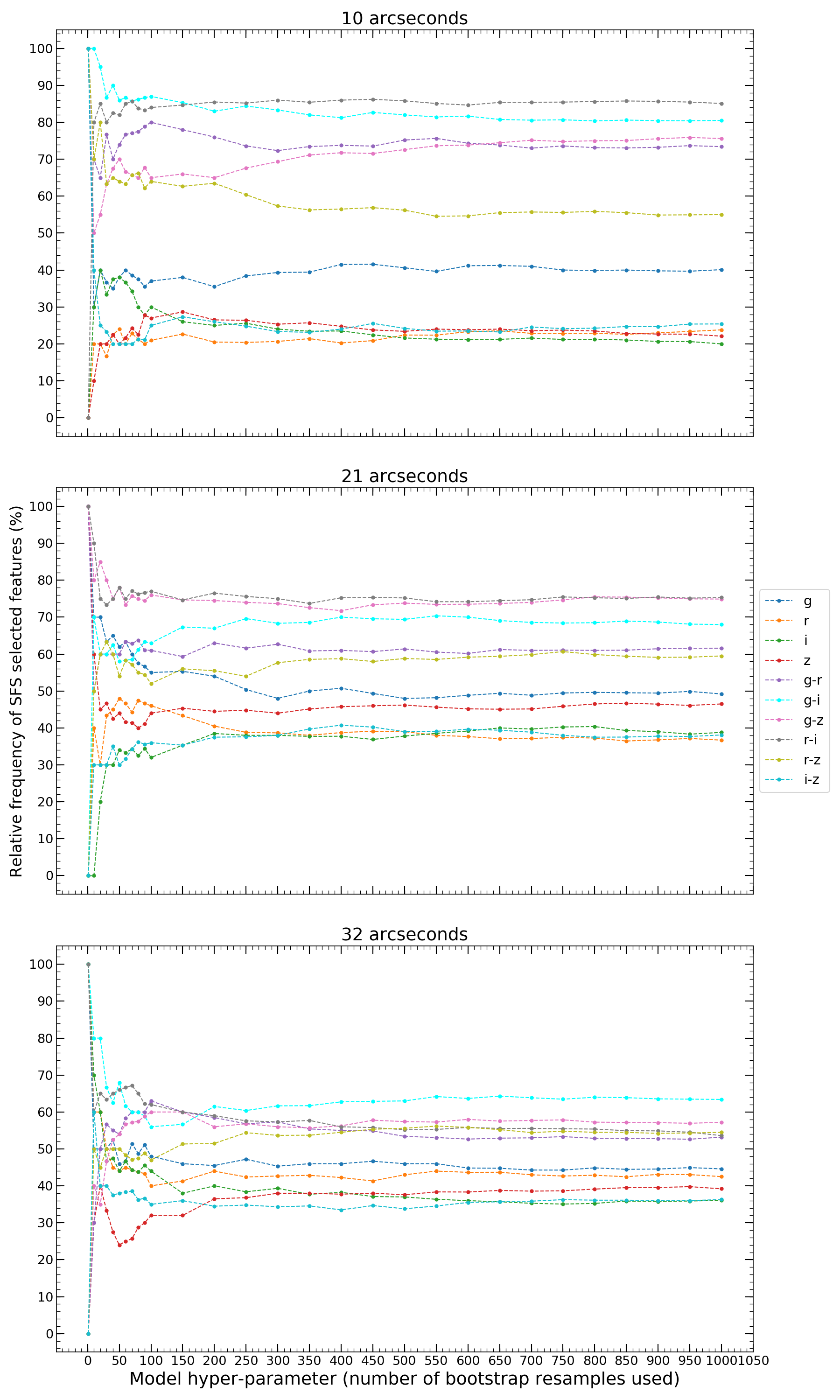}
    \caption{Validation curves from tuning the number of bootstrap resamples hyper-parameter setting, where the relative frequency of features selected by SFS with the MWAR training set is shown for each search radii (10 arcseconds on the top row, 21 arcseconds in the middle row and 32 arcseconds on the bottom row). The individual points display the relative frequency of features selected by SFS across a fixed number of bootstrap resamples with respect to the other hyper-parameter settings of the SRKNN algorithm.}
    \label{fig:validation_curve_bootstraps_features}
\end{figure}

Thirdly, we assess the model performance and behaviour based on the number of bootstrap resamples used for each search radius. Figure \ref{fig:validation_curve_bootstraps} shows that for the MWAR training and validation sets the change in the photometric redshift prediction error steeply decreases when a very small number of bootstrap resamples used but then remains flat as the number of bootstrap resamples increases. This tells us that the number of bootstrap resamples used is not a particularly important hyper-parameter setting to tune as the impact on the model performance for the MWAR training and validation sets is minimal. \cite{number_of_bootstraps} suggests that using fifty to two hundred bootstrap resamples is sufficient to calculate standard errors whereas bootstrap confidence interval estimates require at least one order of magnitude higher computational cost. From which, we decide that using one thousand bootstrap resamples for each search radius would be enough to benefit from bootstrap confidence intervals. We also consider that since SFS would have selected different features for each bootstrap sample, we would not expect all internal KNN models to return predictions after filter magnitude-cuts are applied. Figure \ref{fig:validation_curve_bootstraps_percentages} displays the percentage of clusters returned with full, partial and no bootstrap resamples returned for estimating photometric redshift at each search radius. We find that employing a large number of bootstrap resamples reduces the percentage of clusters returned with no bootstrap resamples. Whilst for clusters with a full set of bootstrap resamples returned, the percentage of clusters returned initially drops but then remains flat as the number of bootstrap resamples increases. Whereas for clusters with partial bootstrap resamples returned, the percentage of clusters returned gradually increases as the number of bootstrap resamples increases. For this work, we prefer to minimise the percentage of clusters returned with no bootstrap resamples, since we want as many clusters as possible to have photometric redshift estimates. In Figure \ref{fig:validation_curve_bootstraps_features} we calculate the relative frequency of features selected by SFS with respect to the number of bootstrap resamples used at each search radius. It can be seen that as the number of bootstrap resamples increases, the spread of the relative frequency amongst the features decreases. From which, we also observe that the features with the highest relative frequency appear to be colours whilst features with the lowest relative frequency are filters. The model has learned that colours are more significant than filters for estimating photometric redshifts of clusters.

\subsection{Model Performance Analysis With Test Sets}
\label{sec:Model_Analysis_with_Test_Set}

We use the WNMR/RNMW test sets to assess the performance of the SRKNN algorithm with the optimal hyper-parameters learned in \S\S\ref{sec:hyper-parameter_analysis} for each search radius. As described earlier in \S\S\ref{sec:prepare_photometry_from_catalogues}, the test sets contain clusters from the WHL12 and redMaPPer cluster catalogues with no corresponding cross-match. A summarised version of the test results can be found in Table \ref{tab:test_set_summary}.

\begin{table*}
    \centering
	\begin{tabular}{|c|c|c|c|c|c|c|}
		\hline
	    Test Set & Search Radius & Optimal Filter Magnitude-Cut & \# Clusters & \# Clusters & \# Clusters & $\widetilde{E_{z}}$ \\ \newline
        & [arcseconds] & [mag] & (total) & (radius) & (tested) &  \\
		\hline
		WNMR & 10 & LM & 9723 & 8844 & 8442 & 0.0106 \\
		WNMR & 21 & LM & 9723 & 9564 & 9057 & 0.013 \\
	    WNMR & 32 & LM-0.5 & 9723 & 9691 & 9057 & 0.014 \\
	    RNMW & 10 & LM & 8646 & 8131 & 7319 & 0.0123 \\
		RNMW & 21 & LM & 8646 & 8577 & 7870 & 0.0156 \\
	    RNMW & 32 & LM-0.5 & 8646 & 8635 & 7416 & 0.0181 \\
		\hline
	\end{tabular}
	\caption{A table displaying the median of photometric redshift prediction errors ($\widetilde{E_{z}}$, where $E_{z} \ = \ |\Delta z|/(1+z)$) across all tested clusters for each test set, search radius and optimal filter magnitude-cut. We also show the total number of clusters in the original full dataset (total), the number of clusters that have galaxies within the specified search radius (radius) and the number of clusters that have galaxies within the specified search radius after filter magnitude-cuts (tested). The values in this table summarise the test results in Figures \ref{fig:whl_pred_vs_actual_50kpc}, \ref{fig:whl_pred_vs_actual_100kpc}, \ref{fig:whl_pred_vs_actual_150kpc}, \ref{fig:redmapper_pred_vs_actual_50kpc}, \ref{fig:redmapper_pred_vs_actual_100kpc} and \ref{fig:redmapper_pred_vs_actual_150kpc}.}
	\label{tab:test_set_summary}
\end{table*}

\begin{figure}
\includegraphics[width=0.49\textwidth]{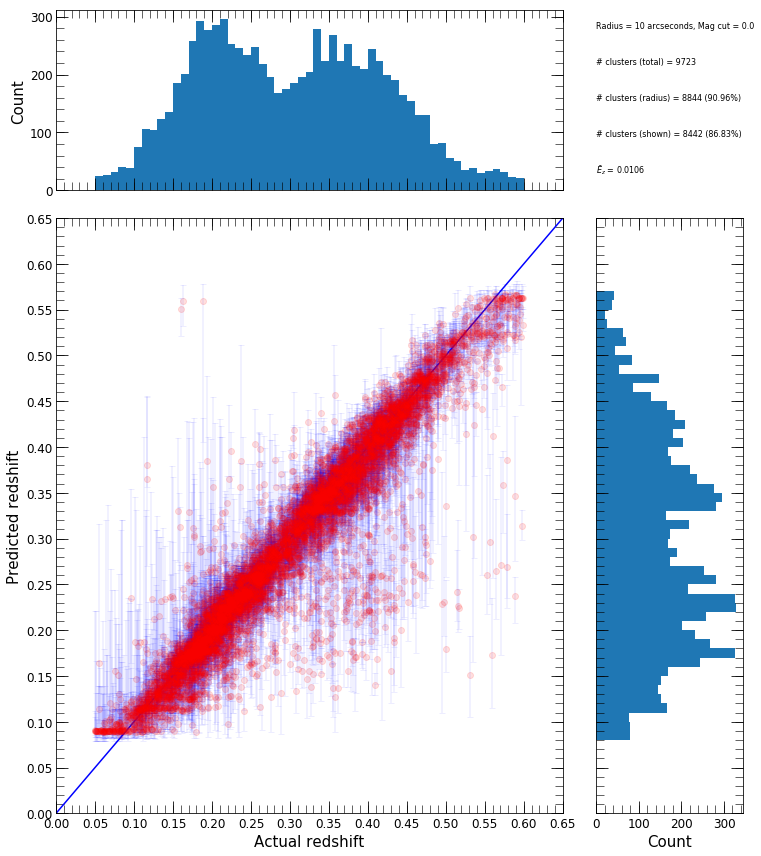}
\includegraphics[width=0.49\textwidth]{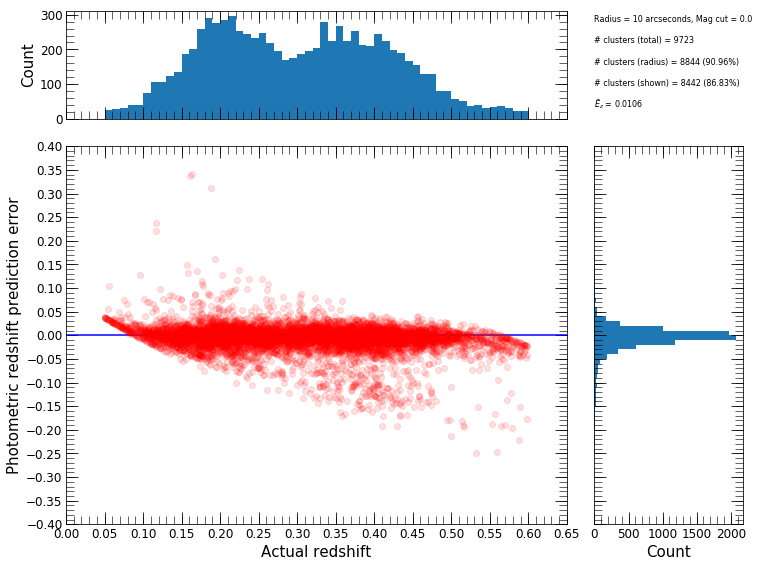}
    \caption{Plots displaying the performance of photometric redshift predictions of clusters for the WNMR test set that had full bootstrap resamples returned within a 10 arcseconds search radius. Top row: Predicted versus `actual' photometric redshift of tested clusters with frequency histograms of the distributions. Bottom row: Non-absolute photometric redshift prediction error versus `actual' redshift of tested clusters with frequency histograms of the distributions. Other: '\# clusters (total)' represents the total number of clusters in the WNMR dataset, '\# clusters (radius)' represents the number of clusters in the WNMR test set that have observed galaxies within a 10 arcseconds search radius, '\# clusters (shown)' represents the number of clusters in the WNMR test set that have observed galaxies within a 10 arcseconds search radius with full bootstrap resamples returned, $\widetilde{E_{z}}$ represents the median of photometric redshift prediction errors across all tested clusters within a 10 arcseconds search radius with partial bootstrap resamples returned.}
    \label{fig:whl_pred_vs_actual_50kpc}
\end{figure}

\begin{figure}
\includegraphics[width=0.49\textwidth]{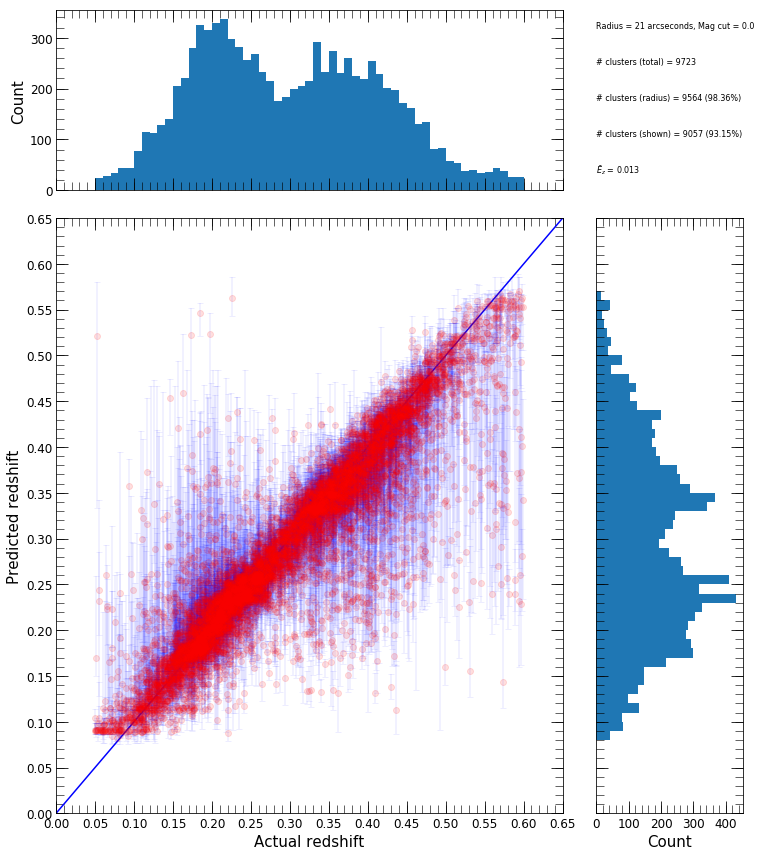}
\includegraphics[width=0.49\textwidth]{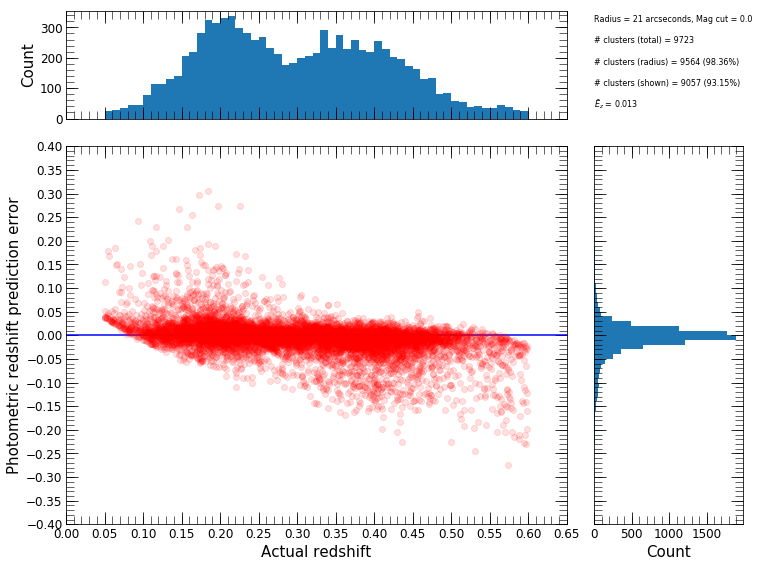}
    \caption{This figure is equivalent to Figure \ref{fig:whl_pred_vs_actual_50kpc} except we examine the performance of photometric redshift predictions of clusters within a 21 arcseconds search radius.}
    \label{fig:whl_pred_vs_actual_100kpc}
\end{figure}

\begin{figure}
\includegraphics[width=0.49\textwidth]{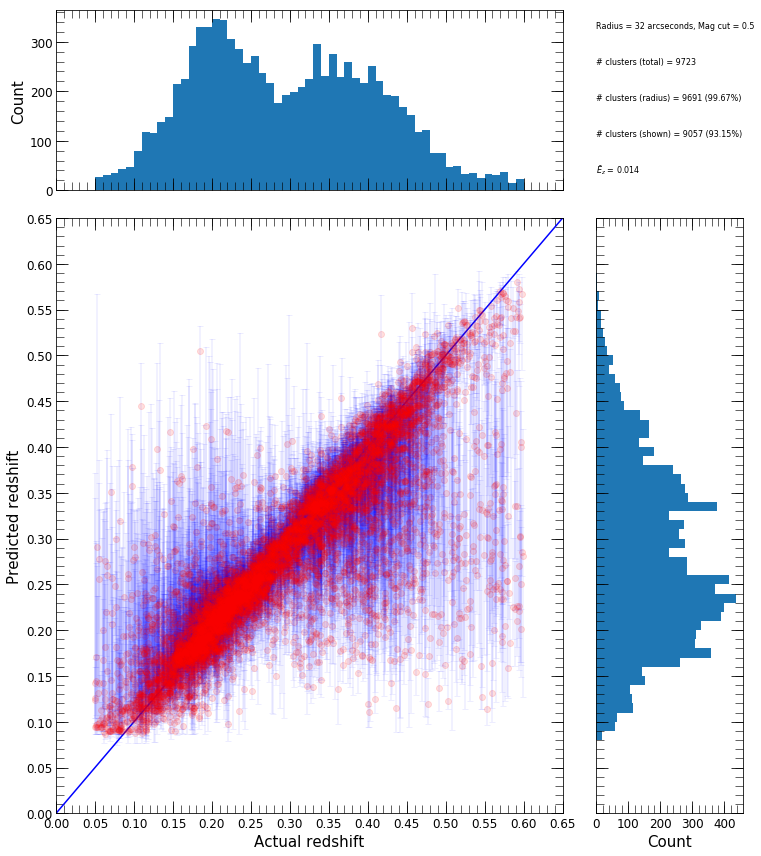}
\includegraphics[width=0.49\textwidth]{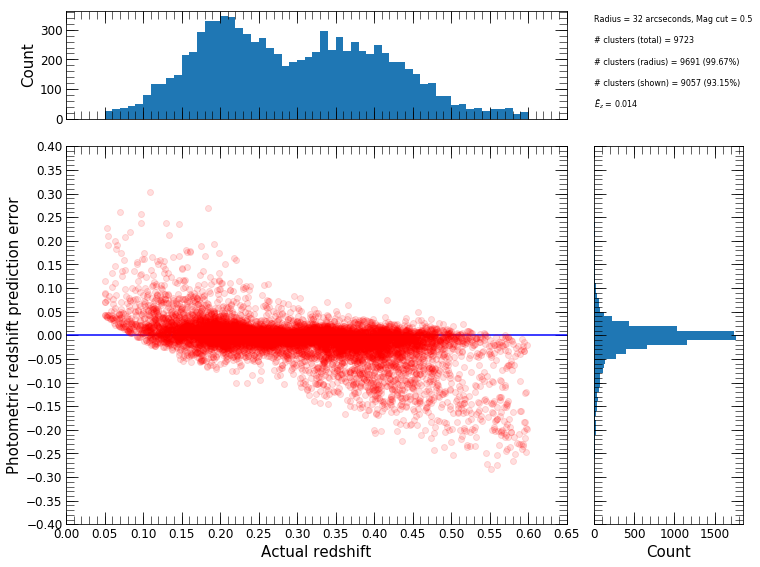}
    \caption{This figure is equivalent to Figure \ref{fig:whl_pred_vs_actual_50kpc} except we examine the performance of photometric redshift predictions of clusters within a 32 arcseconds search radius.}
    \label{fig:whl_pred_vs_actual_150kpc}
\end{figure}

\begin{figure}
\includegraphics[width=0.49\textwidth]{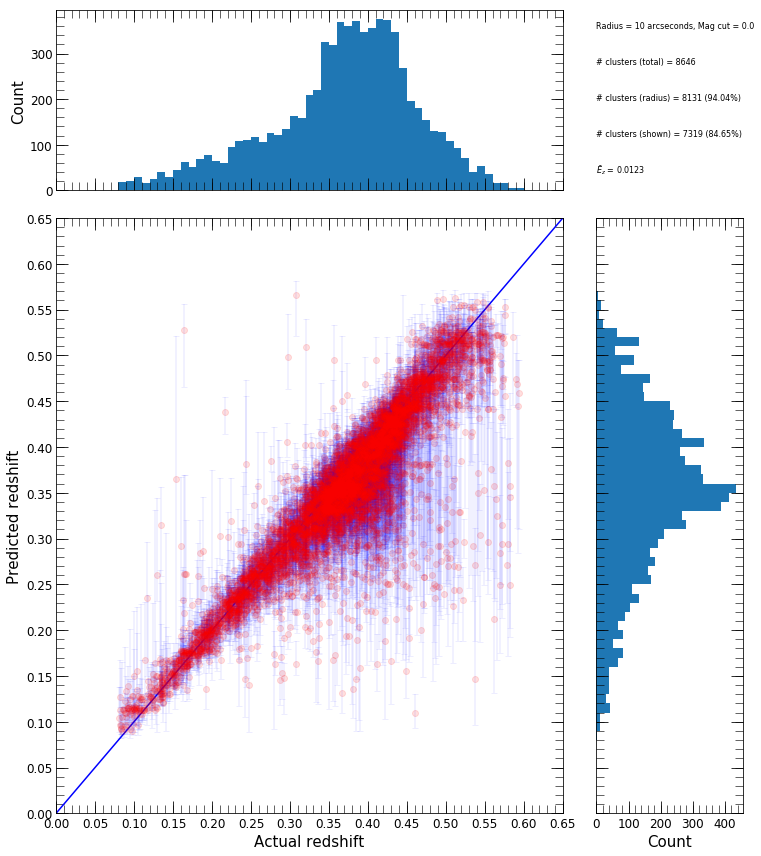}
\includegraphics[width=0.49\textwidth]{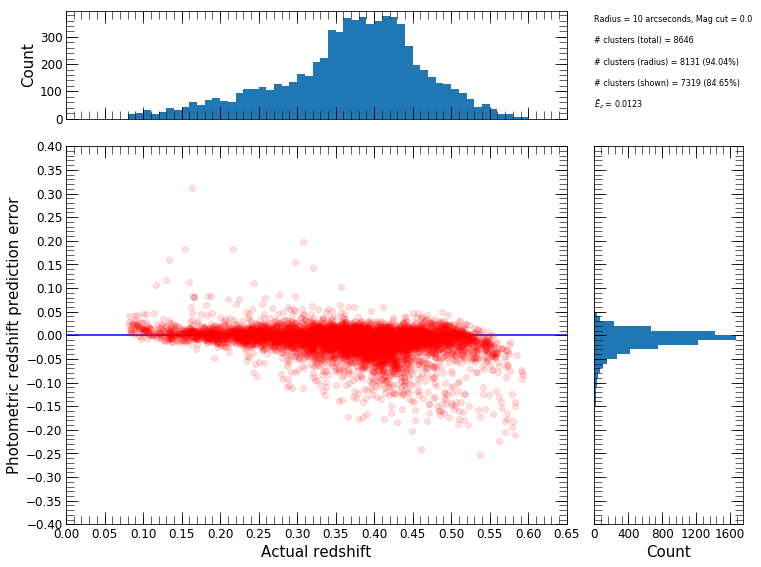}
    \caption{This figure is equivalent to Figure \ref{fig:whl_pred_vs_actual_50kpc} except we examine the performance of photometric redshift predictions of clusters within a 10 arcseconds search radius for the RNMW test set.}
    \label{fig:redmapper_pred_vs_actual_50kpc}
\end{figure}

\begin{figure}
\includegraphics[width=0.49\textwidth]{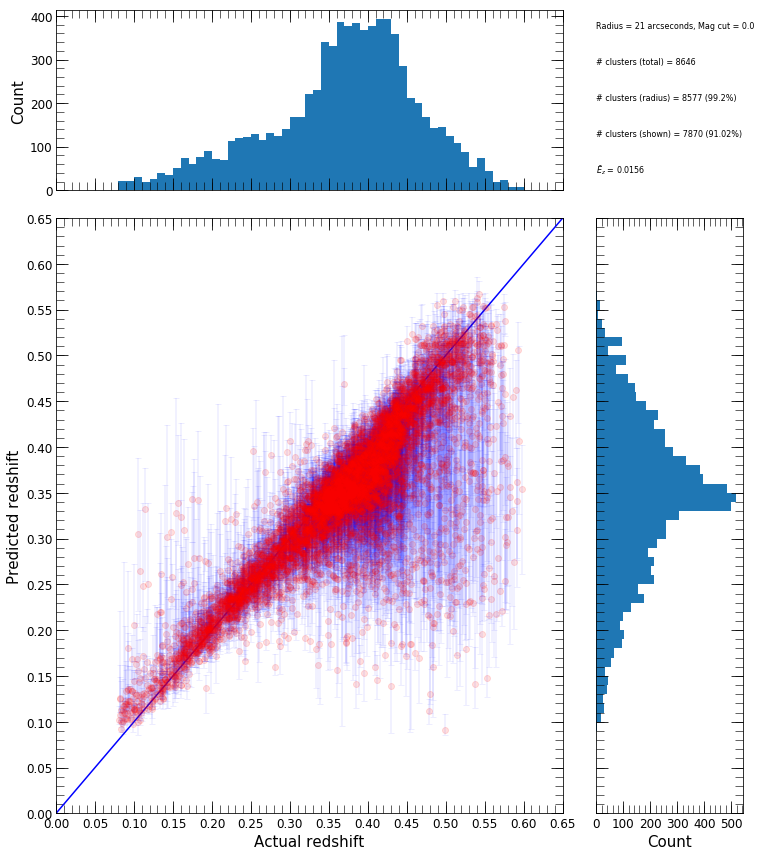}
\includegraphics[width=0.49\textwidth]{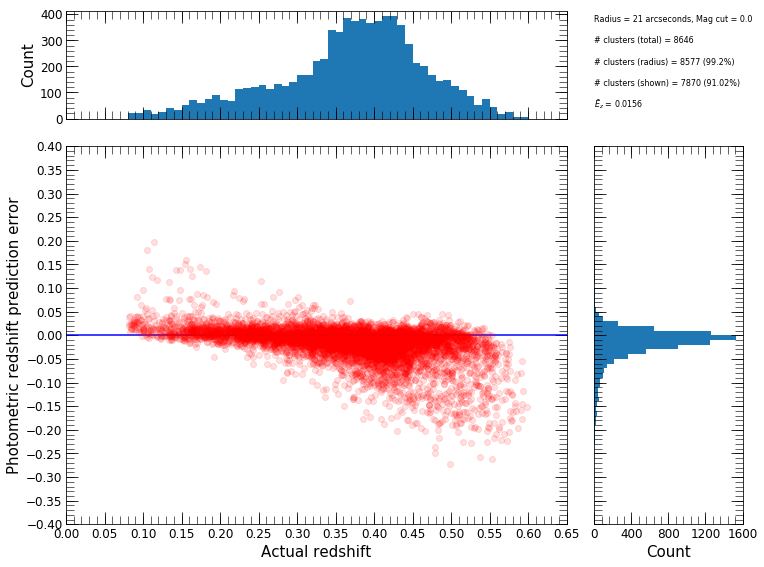}
    \caption{This figure is equivalent to Figure \ref{fig:whl_pred_vs_actual_50kpc} except we examine the performance of photometric redshift predictions of clusters within a 21 arcseconds search radius for the RNMW test set.}
    \label{fig:redmapper_pred_vs_actual_100kpc}
\end{figure}

\begin{figure}
\includegraphics[width=0.49\textwidth]{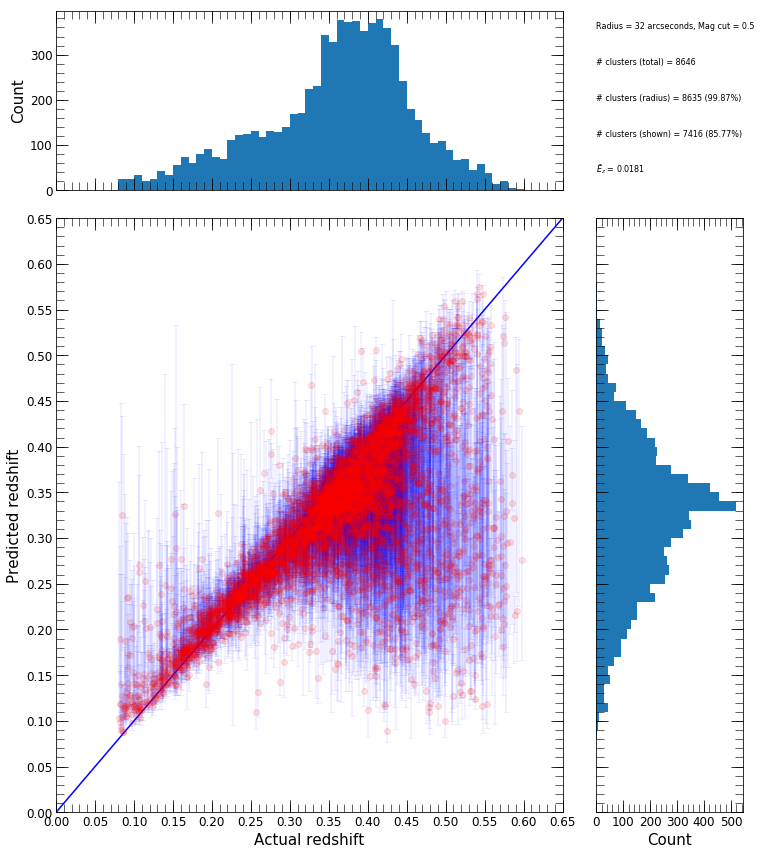}
\includegraphics[width=0.49\textwidth]{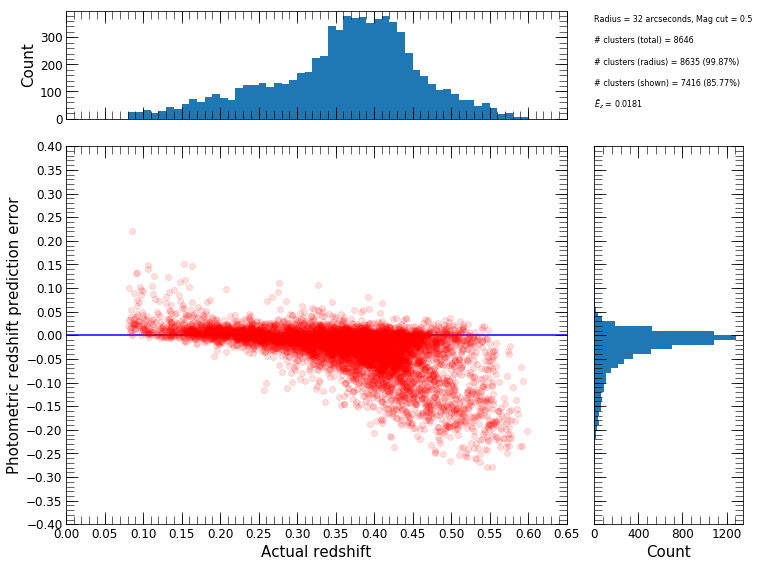}
    \caption{This figure is equivalent to Figure \ref{fig:whl_pred_vs_actual_50kpc} except we examine the performance of photometric redshift predictions of clusters within a 32 arcseconds search radius for the RNMW test set.}
    \label{fig:redmapper_pred_vs_actual_150kpc}
\end{figure}

In Figures \ref{fig:whl_pred_vs_actual_50kpc}, \ref{fig:whl_pred_vs_actual_100kpc}, \ref{fig:whl_pred_vs_actual_150kpc}, \ref{fig:redmapper_pred_vs_actual_50kpc}, \ref{fig:redmapper_pred_vs_actual_100kpc} and \ref{fig:redmapper_pred_vs_actual_150kpc} we compare the known photometric redshifts with the predicted photometric redshifts for clusters in the WNMW/RNMW test sets that had full bootstrap resamples returned by the tuned model. We find that as the search radius increases the median of photometric redshift prediction errors across all tested clusters in both test sets increases as well possibly due to line-of-sight interloping galaxies. From which, in Figures SA4, SA5, SA6, SA7, SA8 and SA9 (available online) we also examine the spatial distribution of several clusters with relatively large photometric redshift prediction errors. We repeatedly observe that if line-of-sight interloping galaxies are present within the search radii of clusters, the resultant model predictions have relatively large photometric redshift prediction errors. Moreover in Figures \ref{fig:whl_pred_vs_actual_50kpc}, \ref{fig:whl_pred_vs_actual_100kpc}, \ref{fig:whl_pred_vs_actual_150kpc}, \ref{fig:redmapper_pred_vs_actual_50kpc}, \ref{fig:redmapper_pred_vs_actual_100kpc} and \ref{fig:redmapper_pred_vs_actual_150kpc} it can be seen that the width of the 95 per cent confidence intervals around predictions decreases as the search radius increases, as shown by wider intervals. This means there is lower precision of the predicted photometric redshift value. Despite this, we find that the tuned model seems to perform well at all redshifts since the majority of cases have relatively low photometric redshift prediction errors for each search radius. Although, we notice that an increasing number of cases have relatively large photometric redshift prediction errors near to the redshift training boundaries of the MWAR training set as the search radius increases. Furthermore, we also examine the performance of the tuned model on clusters in the WNMW/RNMW test sets with only partial bootstrap resamples returned for each search radius. From Figures S2, S3, S4, S5, S6 and S7 (available online) we find that in almost all cases the photometric redshift prediction error is poorly constrained when partial bootstrap resamples are used.

\begin{figure}
\includegraphics[width=0.49\textwidth]{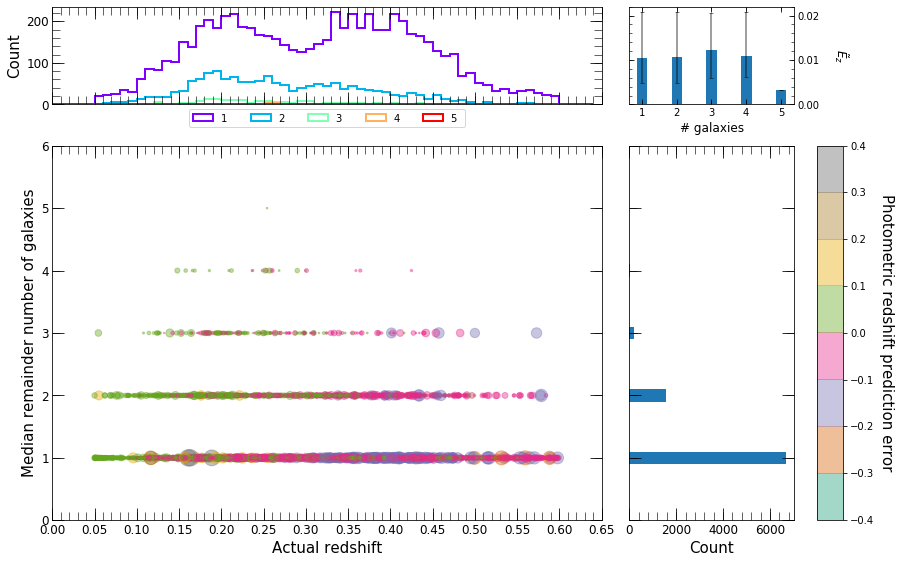}
\includegraphics[width=0.49\textwidth]{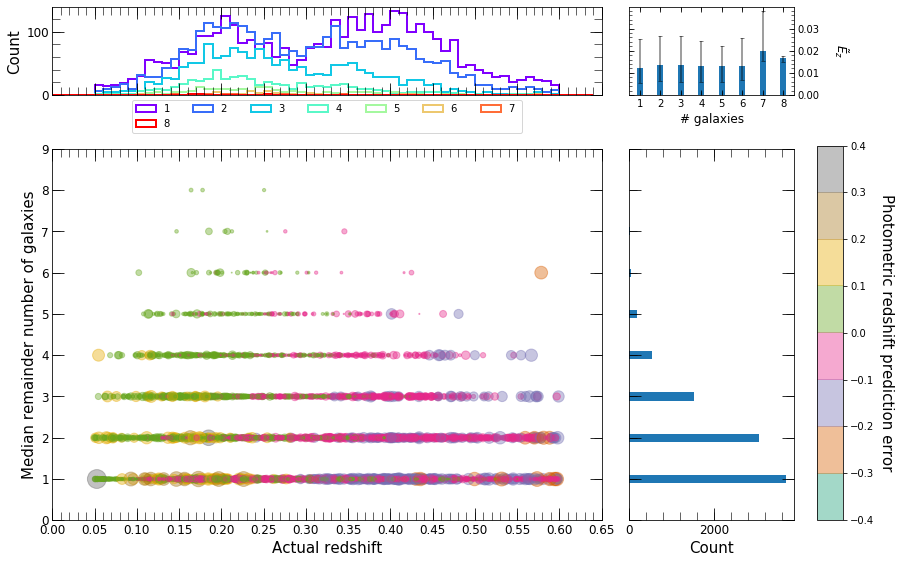}
\includegraphics[width=0.49\textwidth]{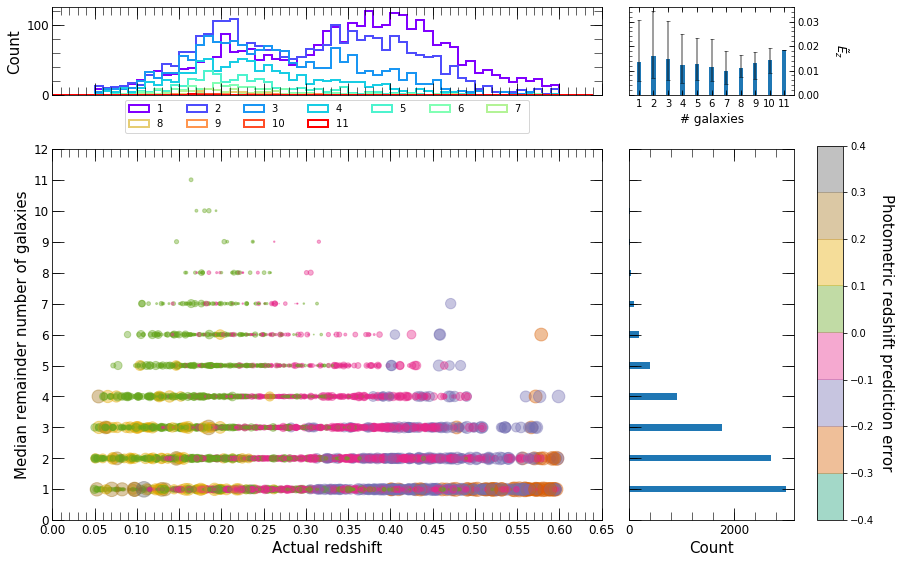}
    \caption{Plots displaying the number of galaxies used in photometric redshift predictions versus `actual' redshift of tested clusters for the WNMR test set, where predictions had full bootstrap resamples returned within a 10 (top row), 21 (middle row) and 32 (bottom row) arcseconds search radius. It should be noted that the size of individual points change in relation to the value of the non-absolute photometric redshift prediction error. Frequency histograms of the distributions are also shown. $\widetilde{E_{z}}$ represents the median of photometric redshift prediction errors across all tested clusters for each number of galaxies bin.}
    \label{fig:whl_number_of_galaxies_all_radius}
\end{figure}

\begin{figure}
\includegraphics[width=0.49\textwidth]{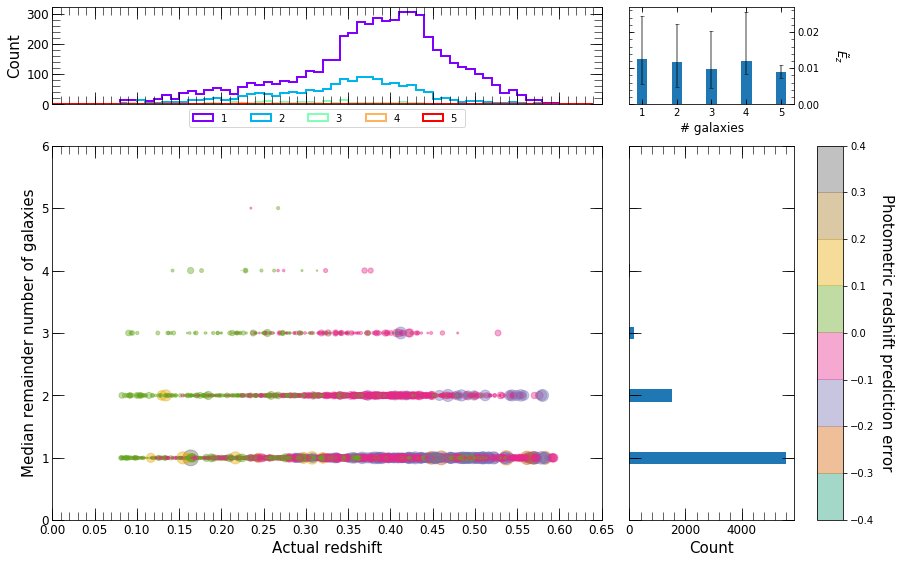}
\includegraphics[width=0.49\textwidth]{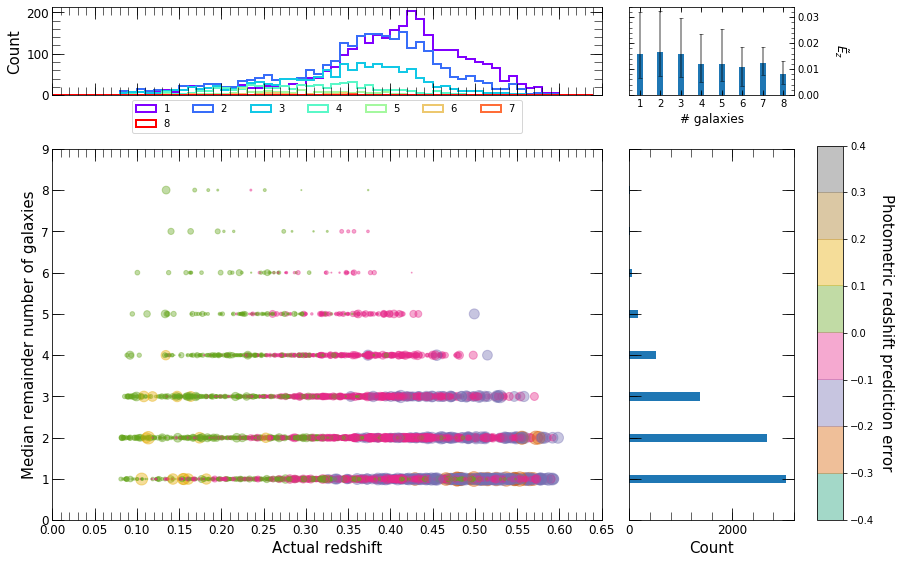}
\includegraphics[width=0.49\textwidth]{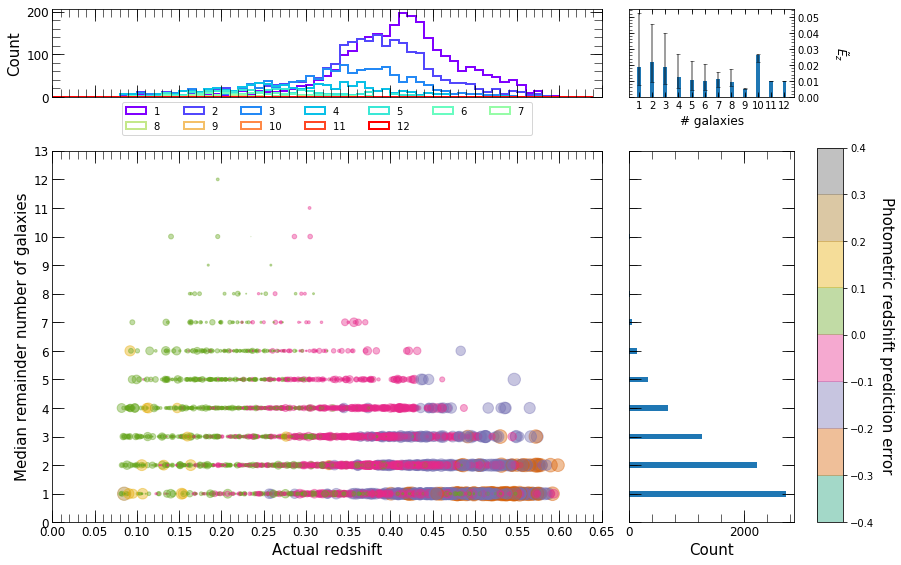}
    \caption{This figure is equivalent to Figure \ref{fig:whl_number_of_galaxies_all_radius} except we examine the number of galaxies used in photometric redshift predictions for the RNMW test set.}
    \label{fig:redmapper_number_of_galaxies_all_radius}
\end{figure}

In Figures \ref{fig:whl_number_of_galaxies_all_radius} and \ref{fig:redmapper_number_of_galaxies_all_radius} we determine the number of galaxies used in photometric redshift predictions of clusters from the WNMR/RNMW test sets that had full bootstrap resamples returned by the tuned model for each search radius. This examines how the tuned model performs with respect to different numbers of galaxies. It can be seen that as the search radius increases the number of galaxies used in photometric redshift predictions increases too. From which, we find that the median of photometric redshift prediction errors across all tested clusters is similar regardless of the number of galaxies used by the tuned model. Although, we notice that clusters with larger numbers of galaxies used for photometric redshift predictions are frequently seen between low and intermediate redshifts with relatively low photometric redshift prediction errors. Whereas clusters at considerably lower and higher redshifts rarely have large numbers of galaxies used for photometric redshift predictions and also have relatively large photometric redshift prediction errors. 

\begin{figure}
\begin{center}
\includegraphics[width=0.4\textwidth, height=0.175\textwidth]{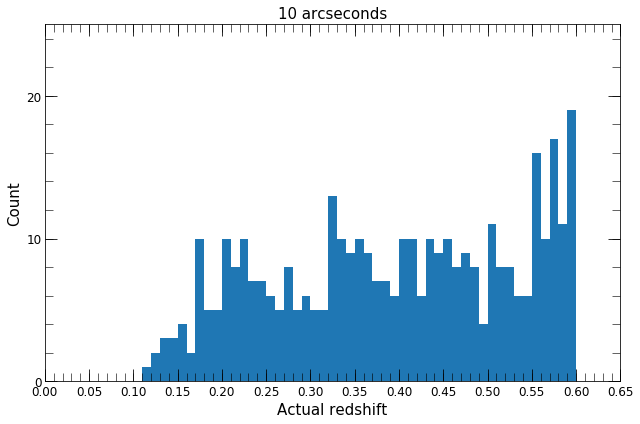}
\includegraphics[width=0.4\textwidth, height=0.175\textwidth]{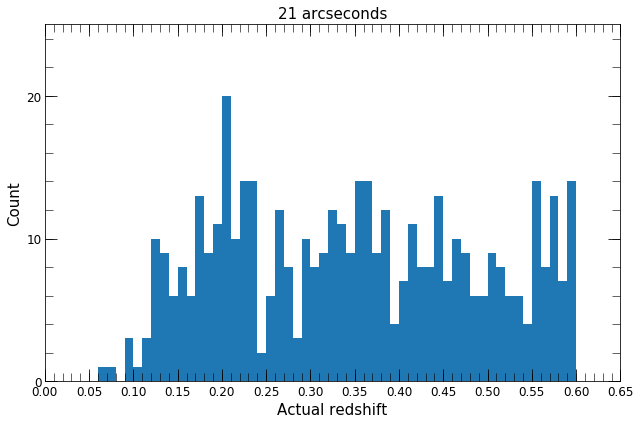}
\includegraphics[width=0.4\textwidth, height=0.175\textwidth]{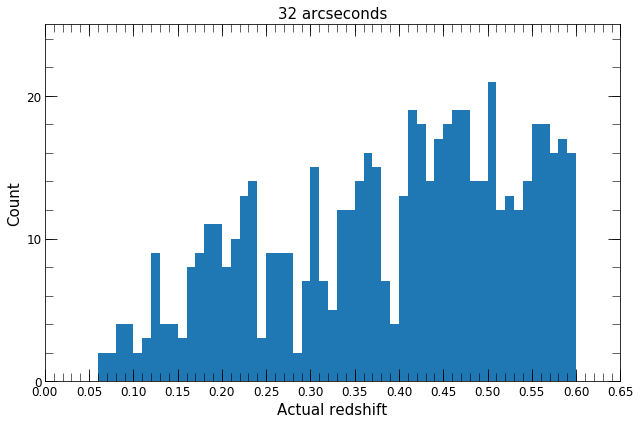}
    \caption{Frequency histograms displaying the `actual' redshift distributions of clusters from the WNMR test set that had no bootstrap resamples returned within a 10 (top row), 21 (middle row) and 32 (bottom row) arcseconds} search radius.
    \label{fig:whl_clusters_with_no_pred}
\end{center}
\end{figure}

\begin{figure}
\begin{center}
\includegraphics[width=0.4\textwidth, height=0.175\textwidth]{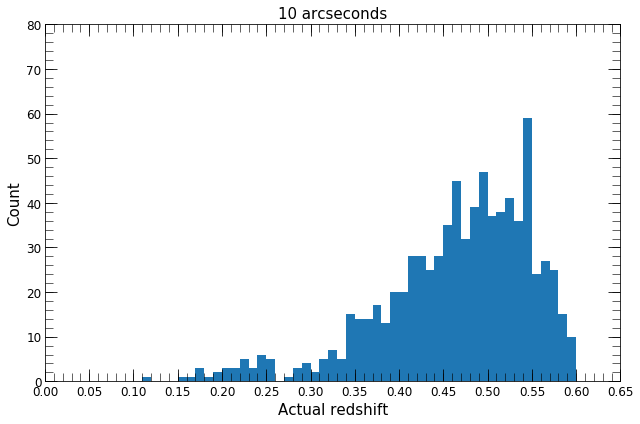}
\includegraphics[width=0.4\textwidth, height=0.175\textwidth]{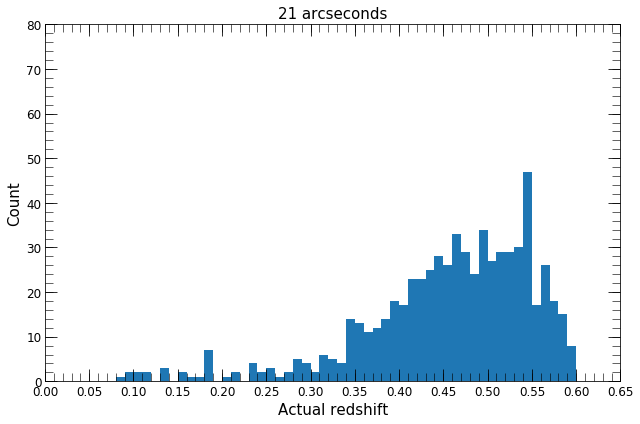}
\includegraphics[width=0.4\textwidth, height=0.175\textwidth]{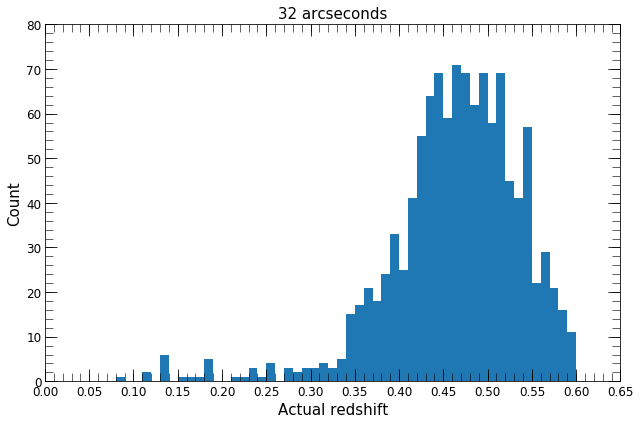}
    \caption{This figure is equivalent to Figure \ref{fig:whl_clusters_with_no_pred} except we examine the `actual' redshift distributions of clusters that had no bootstrap resamples returned for the RNMW test set.}
    \label{fig:redmapper_clusters_with_no_pred}
\end{center}
\end{figure}

In Figures \ref{fig:whl_clusters_with_no_pred} and \ref{fig:redmapper_clusters_with_no_pred} we examine the redshift distribution of clusters from the WNMR/RNMW test sets with no bootstrap resamples returned by the tuned model for each search radius. We observe that the redshift distributions are predominantly skewed towards higher redshifts. This could be due to the galaxies in clusters at higher redshifts having poorer photometric measurements in comparison to the galaxies in clusters at lower redshifts. Although, it should be noted that the redshift distribution for the RNMW dataset itself is also heavily skewed towards higher redshifts.

\subsection{Further Model Testing}
\label{sec:Model_Analysis_with_Additional_Test_Set}

We also test the tuned model on additional clusters that reside in unseen parameter space, such as clusters with low richness and clusters at redshift equal or greater than 0.6. This was to analyse the generalisation of the tuned model, by running it on clusters with properties that it had not been trained for, which are also likely to be encountered in surveys. We apply the same analysis procedure as performed in \S\S\ref{sec:Model_Analysis_with_Test_Set} and provide the full results in the online supplementary material. For this section we will only describe the response of the tuned model with respect to different cluster properties.

In Figures S8, S9 and S10 (available online) we ran the tuned model on clusters with low richness, which have a richness of twenty or fewer member galaxies such that they did not qualify for the MWAR dataset, to obtain photometric redshift predictions that have full bootstrap resamples returned at each search radius. We find that the number of cases with relatively large photometric redshift prediction errors increases as the search radius increases, particularly at higher redshifts. However, we also notice that the median of photometric redshift prediction errors for each search radius remains relatively low when compared to the median of photometric redshift prediction errors for the WNMR/RNMW test sets. Moreover, we observe that the precision of the 95 per cent confidence intervals becomes worse towards the redshift training boundaries when the search radius increases.

In Figures S16, S17 and S18 (available online) we ran the tuned model on clusters at high redshift, which have a redshift beyond the redshift training boundaries such that they did not qualify for the WNMR dataset, to obtain photometric redshift predictions that have full bootstrap resamples returned at each search radius. We immediately notice that the overall accuracy of photometric redshift predictions is low when compared to the other test sets, as the tuned model constantly underestimates the photometric redshifts regardless of the search radius used. We also observe that the precision of the 95 per cent confidence intervals around predictions is poorly constrained, such that it would be difficult to distinguish clusters at high redshift from poorly constrained clusters at intermediate redshift.

In Figures S24, S25 and S26 (available online) we ran the tuned model on clusters at high redshift with low richness, which have a richness of twenty or fewer member galaxies and a redshift beyond the redshift training boundaries such that they did not qualify for the WNMR dataset, to obtain photometric redshift predictions that have full bootstrap resamples returned at each search radius. Similar to the results in Figures S16, S17 and S18 for clusters at high redshift, we find that the overall accuracy of photometric redshift predictions is also low, as the tuned model constantly underestimates the photometric redshifts. In addition, the 95 per cent confidence intervals around predictions are also poorly constrained regardless of the search radius used.

In Figures S32, S33 and S34 (available online) we ran the tuned model on clusters with low richness, which have a richness of twenty or fewer member galaxies such that they did not qualify for the WNMR dataset, to obtain photometric redshift predictions that have full bootstrap resamples returned at each search radius. Similar to the results in Figures S8, S9 and S10 for clusters with low richness, we find that the overall accuracy of the photometric redshift predictions is high, as only a minority of cases have relatively large photometric redshift prediction errors. Although, we also observe that the precision of the 95 per cent confidence intervals becomes worse towards the redshift training boundaries when the search radius increases.

Lastly, we also evaluate the effectiveness from increasing the search radius on the performance of photometric redshift predictions and the number of clusters with full bootstrap resamples returned. For example, if a cluster did not have a photometric redshift estimate with full bootstrap resamples returned within a 10 arcseconds search radius, we would try using a 21 arcseconds search radius instead. From which, if a 21 arcseconds search radius was not sufficient, we would then try using a 32 arcseconds search instead. In Figures S40, S43, S46, S49, S52 and S55 we find that as the search radius increases the overall accuracy of photometric redshift estimates decreases. Although, this can still be beneficial rather than having clusters with no photometric redshift estimates at all. We also observe that as the search radius increases the number of photometric redshift estimates with full bootstrap resamples returned decreases as well. These trends can be seen repeating for all of the test sets.

\section{Discussion}
\label{sec:Discussion}

\subsection{Effectiveness Of Z-Sequence For Photometric Redshift Predictions}
\label{sec:z-sequence_discussion}

In \S\S\ref{sec:Model_Analysis_with_Test_Set} we employ samples from the WHL12 and redMaPPer cluster catalogues to examine the performance of the tuned model. From Figures \ref{fig:whl_pred_vs_actual_50kpc}, \ref{fig:whl_pred_vs_actual_100kpc} and \ref{fig:whl_pred_vs_actual_150kpc} it can be seen that majority of clusters in the WNMR test set are observed at low to intermediate redshifts, whereas from Figures \ref{fig:redmapper_pred_vs_actual_50kpc}, \ref{fig:redmapper_pred_vs_actual_100kpc} and \ref{fig:redmapper_pred_vs_actual_150kpc} it can be seen that majority of clusters in the RNMW test set are observed at intermediate redshift. This tells us that the methods used to estimate photometric redshifts in WHL12 and redMaPPer can significantly influence the resultant redshift distributions. Although, we find that the tuned model does not have much difficulty in working with either of these redshift distributions, as the overall performance of photometric redshift prediction errors for both test sets are similar. From which, we can infer that Z-Sequence can be effectively utilised across a wide range of redshifts if the appropriate training data is available.

For this paper, we assign the photometric redshifts of the WHL12 and redMaPPer cluster catalogues as `actual' redshifts to examine the model performance on a large sample of clusters. Since we aim to minimise data wastage, it is important to try to utilise all available clusters even though not all clusters will have spectroscopic redshifts. We are aware that the `actual' photometric redshifts for clusters in WHL12 and redMaPPer have a scatter of $\sim0.01$ from spectroscopic redshifts. This is similar to the scatter in our photometric redshift prediction errors of $\sim0.01$ from the `actual' photometric redshifts, which suggests that our model is as accurate as it can be based on the data used for training and testing. We expect that our photometric redshift prediction error would decrease if we trained on a large, entirely spectroscopic sample instead as the scatter associated with the photometric redshifts in the WHL12 and redMaPPer catalogues will be removed. In addition, it should be noted that the flaring seen in Figures \ref{fig:redmapper_pred_vs_actual_50kpc}, \ref{fig:redmapper_pred_vs_actual_100kpc} and \ref{fig:redmapper_pred_vs_actual_150kpc} lowers the predicted redshift values between "actual" redshifts of $0.35 \geq z \geq 0.45$ for the RNMW test set. This is due to the flaring originating from redMaPPer itself and not from our algorithm, as it also occurs in Figure 7 of \cite{redMapper}.

In \S\S\ref{sec:Model_Analysis_with_Additional_Test_Set} we test the tuned model on clusters with unseen properties. We find that the tuned model performs well on clusters in similar parameter space to the MWAR training set and it also performs well on clusters of all richnesses within the redshift training boundaries. However, the tuned model performs poorly on clusters beyond the redshift training boundaries. This tells us that the performance of the tuned model is more dependent on the redshift of the cluster than the richness of the cluster. The tuned model is only effective on clusters at the redshift range it was trained for since we are limited to the redshift range of the majority of clusters available in SDSS. In addition, we observe an apparent feature seen at the lower and upper boundaries for predicted photometric redshifts in Figures \ref{fig:whl_pred_vs_actual_50kpc}, \ref{fig:whl_pred_vs_actual_100kpc} and \ref{fig:whl_pred_vs_actual_150kpc}. We believe the cause of the apparent feature is due the nature of the machine learning algorithm itself. This is because the k-nearest neighbours algorithm calculates its prediction from the labels of the nearest neighbour examples in the training set when given an input data point, where the photometric redshift limits of the MWAR dataset is $0.0698 \leq \textit{z} \leq 0.5986$ whilst the WNMR dataset is $0.05 \leq \textit{z} \leq 0.599$. This means that all photometric redshift predictions are bounded within the photometric redshift training range, such that clusters with `actual' redshifts outside the boundaries could end up as part of the apparent feature. This explains why we do not observe the apparent feature in Figures \ref{fig:redmapper_pred_vs_actual_50kpc}, \ref{fig:redmapper_pred_vs_actual_100kpc} and \ref{fig:redmapper_pred_vs_actual_150kpc} as the photometric redshift limits of the RNMW dataset is $0.0811 \leq \textit{z} \leq 0.5983$. As a further demonstration of the success of our algorithm, we note that the WNMR and RNMW test sets consist of clusters found in one catalogue and not the other. This could mean that these clusters are more difficult to detect and therefore potentially harder to assign a redshift value via other photometric redshift prediction methods, whereas our algorithm can estimate redshifts for the majority of these clusters. It should also be noted that the observed magnitude errors for all SDSS filters increases with redshift, as seen in Figure SA10 (available online). This means it would be difficult for any empirical algorithm to make accurate photometric estimates in the high redshift regime. However, we expect our model would be successful at estimating photometric redshifts for high redshift clusters if trained on imaging surveys such as LSST or Euclid, which will have greater photometric depths to increase the redshift limits of cluster detection when compared with SDSS. 

We notice in Table \ref{tab:test_set_summary} that the median value of $|\Delta z|/(1+z)$ increases for the WNMR and RNMW test sets by 32 per cent and 47 per cent respectively when the search radius is enlarged from 10 arcseconds to 32 arcseconds. This can also be seen in \S\S\ref{sec:Model_Analysis_with_Additional_Test_Set} where the number of cases with accurate photometric redshift estimates decreases as the search radius increases, as a larger search radius is more likely to include interlopers. From Figures SA6, SA7 and SA9 (available online), we find that interlopers are evident in contaminating estimates with relatively large photometric redshift prediction errors if they appear in the test set. Whilst Figures SA4, SA5 and SA8 (available online) indicate that interlopers are also somewhat present within the training set itself, as we find that some model predictions for clusters with no obvious interlopers in the test set still have relatively large photometric redshift prediction errors. Subsequently, we aim to further improve the accuracy of the Z-Sequence model in future work by developing new strategies to constrain interlopers, such as with unsupervised machine learning techniques that identifies the presence of line-of-sight interloping galaxies and multiple projected line-of-sight clusters. This new method can be employed as an additional pre-processing tool to accompany the Z-Sequence model. From which, we could increase the size of the search radius once the obvious interlopers are removed and examine whether the photometric redshift prediction accuracy significantly improves if more cluster members are included. In addition, Figures SA6 and SA9 (available online) show that filter magnitude-cuts are also partially responsible for estimates with relatively large photometric redshift prediction errors, as we find that all of the galaxy members in some cluster cores are removed from model predictions due to poor photometry measurements. Furthermore, we notice in Figure SA7 (available online) that the 95 per cent confidence interval for the photometric redshift estimate involving the interloper becomes considerably wider in comparison to the photometric redshift estimates without the interloper. This shows that the bootstrap confidence intervals could indicate whether interlopers are involved in the model prediction. Although, it should be noted that Figures \ref{fig:whl_number_of_galaxies_all_radius} and \ref{fig:redmapper_number_of_galaxies_all_radius} show that the majority of the model predictions seem to employ relatively few galaxies for each search radii, such that it would be difficult to constrain interlopers in most instances. Moreover, by comparing the number of clusters that have photometric redshift estimates with full bootstrap resamples returned exclusively within each of the 10, 21, 32 arcseconds search radii (see Figures S40, S43, S46, S49, S52 and S55 [available online]), we discover that the majority of cases are actually within the 10 arcseconds search radius whereas only a minority of cases require an increase in the search radius to 21 and 32 arcseconds. This suggests that if we were to retrain the model on different surveys, we could consider not needing to employ multiple large search radii as the computational cost for training the model could outweigh the benefits gained.

It is worth noting that our approach results in photometric redshift predictions with full, partial and no bootstrap resamples returned. This is primarily due to the use of filter magnitude-cuts in each internal KNN model, which excludes galaxies with poorer photometric measurements from the cluster before any predictions are made. Although, we observe in \S\S\ref{sec:feature_selection_analysis} that applying filter magnitude-cuts can improve the overall accuracy of photometric redshift estimates. From which, we find that photometric redshift predictions with full bootstrap resamples returned are fairly accurate, as seen in \S\S\ref{sec:Model_Analysis_with_Test_Set}. However, it can also be seen that photometric redshift predictions with partial bootstrap resamples returned have low accuracy. This could be caused by the remaining bootstrap resamples not utilising strong predictive features. Subsequently, we advise that future photometric redshift estimates with partial bootstrap resamples returned should be flagged and used cautiously.

\subsection{Practicality Of The Machine Learning Techniques Used In This Paper}
\label{sec:machine_learning_discussion}

For this paper, we are aware that the KNN algorithm can suffer from a dimensionality effect known as the `curse of dimensionality' \citep{curse_of_dimensionality}. This can cause training samples to be disproportionately represented and sparsely distributed in high dimensional feature space, especially when the number of input features is greater than the number of training samples. As a consequence, this restricts the performance of machine learning algorithms due to the high complexity learning involved. There are several approaches that can be used to limit the impact of this dimensionality effect, which include feature selection techniques (e.g. Sequential Feature Selection [\citealt{feature_selection}], Chi-Squared Test [\citealt{chi-squared_test}], Fisher Score [\citealt{fisher_score}]) and feature extraction techniques (e.g. Principal Component Analysis [\citealt{principal_component_analysis_0}; \citealt{principal_component_analysis_1}], Independent Component Analysis [\citealt{independent_component_analysis_0}; \citealt{independent_component_analysis_1}], Partial Least Squares Regression [\citealt{partial_least_squares_regression_0}; \citealt{partial_least_squares_regression_1}]). These techniques promote useful features and ignore redundant features to subsequently constrain the dimensionality of the feature space. For a classification scenario, \cite{curse_of_dimensionality_solution} suggests that if the number of input features is not too large, such as between five to ten, then at least between fifty to one hundred corresponding training samples are required per class to minimise the `curse of dimensionality'. In our case, we ensure that the MWAR training set has a sufficient number of observations in the majority of redshift bins, as seen in Figure SA2 (available online). In addition, we prefer to use a feature selection method that employs features which maximise prediction accuracy rather than a feature extraction method that projects statistically significant features into a reduced feature space. 

The most commonly used sequential feature selection strategies are Sequential Forward Selection (SFS) and Sequential Backward Selection (SBS). These methods are designed to be computationally efficient by searching through fewer combinations of feature space to provide a quasi-optimal solution rather than a global optimal solution. As described earlier in \S\S\ref{sec:feature_selection_process}, SFS iteratively adds features to an empty feature subset in a forward manner whilst SBS iteratively eliminates features from a full feature subset in a backward manner (\citealt{sequential_feature_selection_algorithms}). This means that SBS will examine more high dimensional combinations of features compared with SFS, which could increase prediction accuracy but at a much higher computational cost. Nonetheless, we decide that SBS is not compatible for this work since the 95 per cent cluster retainment threshold would be immediately bypassed if all features are used at the same time, as seen in Figure \ref{fig:filter_mag_cut}. Although, we could consider SBS as an alternative feature selection strategy in imaging surveys that have greater filter sensitivity than SDSS. We also compare the performance between SFS and manual feature selection. From comparing Figures \ref{fig:feature_frequency_and_predictions} and S1 (available online), we find that SFS selected features consistently perform better than the manually selected features for the KNN algorithm. This means that SFS is more precise than manual feature selection at taking into account minor details in the datasets. From which, we decide that SFS has better synergy for working with bootstrap resamples in the SRKNN algorithm. It should be noted that we also randomly initialise the input features to the SRKNN algorithm as an additional starting step to SFS to reduce the impact from strong collinear features (see Figures SA11 and SA12 [available online]) during the feature selection process. Furthermore, in Figure \ref{fig:validation_curve_bootstraps_features} it can be seen that using a large number of bootstrap resamples for the SRKNN algorithm improves the stability for the relative frequency of SFS selected features. This is in contrast to using an individual algorithm such as the KNN algorithm (see Table \ref{tab:feature_frequency}) or using just one bootstrap resample in the SRKNN algorithm. This tells us that the SRKNN algorithm with a large number of bootstrap resamples is able to cope with minor changes to the training set, which would otherwise result in completely different features being used by the model. 

The bias-variance tradeoff describes how generalised a supervised machine learning algorithm is at learning a target function \citep{bias_variance_tradeoff}. If an algorithm is highly dependent on the training dataset during learning, it will perform poorly on new data. This results in many predictions with high variance and low bias. On the other hand, if an algorithm makes a lot of assumptions from the training dataset during learning, it will reduce the predictive power of the algorithm. This results in many predictions with high bias and low variance. For example, the bias-variance tradeoff for the KNN algorithm varies depending on the number of nearest neighbours used, where using low values for the number of nearest neighbours can induce overfitting whilst using high values for the number of nearest neighbours can induce underfitting (\citealt{knn_bias_variance_tradeoff_0}; \citealt{knn_bias_variance_tradeoff_1}). For the SRKNN algorithm, we examine a wide range of number of nearest neighbours from 1 to 25 but this range could be extended with increased computation in future work to explore a larger number of nearest neighbours. In \S\S\ref{sec:hyper-parameter_analysis} we had chosen a value for the number of nearest neighbours that shows no obvious indications of overfitting or underfitting. It is also known that ensemble algorithms can intrinsically reduce the overall variance of predictions for a model by averaging estimates from multiple models that individually have high variance predictions (e.g. \citealt{ensemble_variance_reduction}). This effect can be observed in the random forest (RF, \citealt{random_forest}) algorithm, which is an ensemble that averages the estimates from multiple decision trees (DT, \citealt{decision_trees_0}; \citealt{decision_trees_1}). 

The main difference between the SRKNN and RF algorithms is the choice of internal model, such that each ensemble is better suited for different applications. The KNN algorithm utilises instance-based learning \citep{instance-based_learning}, which means it has no learnable parameters. Whilst the DT algorithm utilises partition-based learning \citep{partition-based_learning}, which means it learns optimal splitting parameters. It should be noted that the KNN algorithm can support a similar partition strategy to the DT algorithm by utilising K-Dimensional Tree \citep{kd_tree} or Ball Tree search \citep{ball_tree}. Generally, the KNN algorithm provides higher flexibility for evaluating complex patterns whereas the DT algorithm has greater interpretability for understanding underlying decisions \citep{knn_vs_decision_tree}. In Figures SA13, SA14, SA15, SA16, SA17 and SA18 (available online) we use the t-Distributed Stochastic Neighbour Embedding (t-SNE, \citealt{t-sne_algorithm}) algorithm to visualise how the feature space of the MWAR training set appears in two-dimensional space with and without feature scaling applied for each search radius. We observe that galaxies with similar photometric redshifts are somewhat clustered to form smooth transitions from low to intermediate redshifts when feature scaling is applied. Moreover, we also observe that galaxies with similar photometric redshifts are considerably dispersed across feature space when feature scaling is not applied. Nevertheless, the structure of these feature spaces would be difficult for the DT algorithm to apply partitions, whilst the KNN algorithm is better suited to work with these smooth transitions, regardless of whether feature scaling is applied. From which, the SRKNN algorithm would also be more applicable at handling photometry data to estimate photometric redshifts than the RF algorithm.

There are numerous hyper-parameter setting optimisation strategies available for machine learning algorithms that are suited for different situations. The most commonly used strategies are grid search, random search \citep{grid_and_random_search} and Bayesian optimisation \citep{hyperparameter_optimisation}. These strategies require the user to define a range of hyper-parameter setting values that will be explored. The simplest approach is grid search, which evaluates all combinations of hyper-parameter settings but this approach can incur high computational cost. Whereas random search can be computationally cheaper, as it iteratively examines random combinations of hyper-parameter settings to compute an approximate solution. For machine learning algorithms with relatively few hyper-parameter settings, such as linear regression, grid search is more preferable to determine optimal hyper-parameter settings. However, as the number of hyper-parameter settings increases, it becomes computationally favourable to apply random search. Alternatively, if the number of hyper-parameter settings is relatively large, such as neural networks, then it is applicable to employ Bayesian optimisation. This uses Bayes theorem (\citealt{bayes_theorem_0}; \citealt{bayes_theorem_1}) to generate probability estimates of the optimal hyper-parameter settings, which involves incorporating prior assumptions of the hyper-parameter settings and iteratively updating a probabilistic distribution of the search space. This means that Bayesian optimisation can minimise the number of hyper-parameter setting combinations that need to be tested. Although in this work, we decide that it is appropriate to utilise grid search to determine the optimal hyper-parameter settings, since the number of hyper-parameter settings for the SRKNN algorithm is relatively low. 

We are also aware that the accuracy of photometric redshift estimates has a dependency on the accuracy of the cluster finder used to locate the cluster. For this work, we treat all input data points in CMS with uniform distance weighting. This means that all input data points are not influenced by the distance to the training set data points. However, this may reduce the accuracy of photometric redshift estimates in regions of the sky that have many line-of-sight interloping galaxies since the cluster finder would be unable to cleanly define the cluster core, where the red sequence is most well-defined. To limit the dependency on the cluster finder, we could consider simple non-uniform weighting strategies for the SRKNN algorithm such as inverse distance weighting \citep{inverse_distance_weightings}. This computes weights based on the distance of the input data points to the training set data points, where the significance of the training set data points decreases as the distance increases. The reason we do not utilise this approach is due to the fact that it is also highly susceptible to noise in the training set. Although, in future work we could consider inverse distance weighting as an alternative, if we can further constrain line-of-sight interloping galaxies within the training set. In addition, the reason we do not utilise photometric redshift estimates of individual galaxies determined by SDSS itself is due to the fact that our method allows us to operate in situations where no photometric redshifts of individual galaxies are available.  

In k-fold cross validation the dataset is partitioned into `k' number of folds, whilst in hold-out validation the dataset is split into distinct sets. For k-fold cross validation five or ten `k' folds is commonly employed, whereas for hold-out validation a seventy/thirty or eighty/twenty percentage split of the dataset is typically applied. Each approach is suited for different circumstances to balance between computational cost and bias-variance sample misrepresentation tradeoff \citep{ten_fold_cv}. This means that k-fold cross validation benefits from a low variance evaluation at a high computational cost. Whereas hold-out validation produces a high variance evaluation but for a low computational cost. In this work, we decide that ten-fold cross validation is appropriate for feature selection and filter magnitude-cut analysis of the KNN algorithm, as the KNN algorithm has moderate computational training cost requirements. On the other hand, the SRKNN algorithm has higher computational training cost requirements especially when a large number of bootstrap resamples is used. From which, we decide that hold-out validation is more preferable for hyper-parameter tuning of the SRKNN algorithm. However, with increased computation we could consider using k-fold cross validation for hyper-parameter tuning in future work. 

\section{Conclusion}
\label{sec:Conclusion}

We present Z-Sequence, an empirical model that is composed of an ensemble of the k-nearest neighbours algorithm, known as the sequential random k-nearest neighbours algorithm. The model makes use of photometry data from observed galaxies within a specified search radius to estimate photometric redshifts of clusters. In this proof-of-concept study, we assembled training sets with cross-matched clusters detected in the Sloan Digital Sky Survey by the WHL12 and redMaPPer cluster catalogues, as using cross-matched clusters reduced the likelihood of having false detections in the training set. Whilst clusters that were not cross-matched were used to test the performance of the model. We demonstrated that employing an automated feature selection strategy, known as sequential forward selection, is effective at identifying predictive features from an initial set of features (i.e. filters and colours). We have shown that applying filter magnitude-cuts to the photometry data improved the overall accuracy of photometric redshift estimates, as this excluded galaxies with poor photometric measurements from model predictions. We examined the behaviour of each hyper-parameter setting for the SRKNN algorithm to understand how varying them affected model performance and generalisation. From which, we found that the choice of the number of nearest neighbours had the biggest impact, the choice of the number of initialised random features had moderate impact and the choice of the number of bootstrap resamples used had the least impact. The optimal values for each hyper-parameter setting were subsequently chosen for model testing. Our results showed that the tuned model performed well on clusters that were within the same redshift range (i.e. low and intermediate redshift) as the clusters in the training set and we also demonstrated that the tuned model is effective on clusters of all richnesses that were within the redshift training boundaries. We have shown the photometric redshift prediction error of Z-Sequence via the median value of $|\Delta z|/(1+z)$ on the WHL12 test samples (across a photometric redshift range of $0.05 \leq \textit{z} \leq 0.599$) to be 0.0106 and on the redMaPPer test samples (across a photometric redshift range of $0.081 \leq \textit{z} \leq 0.598$) to be 0.0123 within a 10 arcseconds search radius, where the photometric redshift prediction error for both test samples increased by 32 per cent and 47 per cent respectively when the search radius is enlarged to 32 arcseconds. In future work, we aim to apply our technique to imaging surveys as a tool to approximate redshifts for many clusters, such as LSST \citep{lsst_survey}, Euclid Survey (\citealt{euclid_survey_0}; \citealt{euclid_survey_1}), Wide Field Instrument High Latitude Survey \citep{romanspacetelescope_survey_0}, Hyper Suprime-Cam Subaru Strategic Survey (\citealt{hsc_survey_0}; \citealt{hsc_survey_1}; \citealt{hsc_survey_2}), Dark Energy Survey (\citealt{dark_energy_survey_0}; \citealt{dark_energy_survey_1}) and XMM Cluster Survey \citep{xcs_survey_0}. It should be noted that our approach has no prerequisites which means that it is fully data driven. This is beneficial for photometric redshift estimation since Z-Sequence can be adapted to any imaging survey and trained on galaxy photometry data from known cluster positions in existing cluster catalogues. To prepare for upcoming surveys, we intend to run Z-Sequence as a complementary tool to our own DEEP-CEE \citep{deep_cee} cluster finder to examine the entirety of the SDSS sky coverage in a preliminary data pipeline, where clusters detected directly from the astronomical images would be accompanied with estimated photometric redshifts.

\section*{Acknowledgements}

We would like to thank the anonymous referee for their thorough feedback which has improved the clarity of our paper.

We gratefully acknowledge the support from the Science and Technologies Facilities Council studentship funding and from the High End Computing facility at Lancaster University to perform extensive computations.

We would also like to thank the developers of Vizier \citep{vizier}, Risa Wechsler at Stanford University, TOPCAT \citep{TOPCAT}, James Schombert at the University of Oregon, Edward L. Wright at the University of California, Los Angeles \citep{cosmo_calc} and Scikit-Learn \citep{scikit-learn} for allowing the open distribution and free usage of their software for research.

Funding for SDSS-III has been provided by the Alfred P. Sloan Foundation, the Participating Institutions, the National Science Foundation, and the U.S. Department of Energy Office of Science. The SDSS-III web site is \url{http://www.sdss3.org/}.

SDSS-III is managed by the Astrophysical Research Consortium for the Participating Institutions of the SDSS-III Collaboration including the University of Arizona, the Brazilian Participation Group, Brookhaven National Laboratory, Carnegie Mellon University, University of Florida, the French Participation Group, the German Participation Group, Harvard University, the Instituto de Astrofisica de Canarias, the Michigan State/Notre Dame/JINA Participation Group, Johns Hopkins University, Lawrence Berkeley National Laboratory, Max Planck Institute for Astrophysics, Max Planck Institute for Extraterrestrial Physics, New Mexico State University, New York University, Ohio State University, Pennsylvania State University, University of Portsmouth, Princeton University, the Spanish Participation Group, University of Tokyo, University of Utah, Vanderbilt University, University of Virginia, University of Washington, and Yale University.

\section*{Data Availability}

The photometry data used in this article is publicly available from the Sloan Digital Sky Survey at \url{https://vizier.u-strasbg.fr/viz-bin/VizieR?-source=V/139}. The WHL12 and redMaPPer v6.3 cluster catalogues can also be found in the public domain at \url{http://vizier.u-strasbg.fr/viz-bin/VizieR?-source=J/ApJS/199/34} and \url{http://risa.stanford.edu/redmapper/}.








\bsp	
\label{lastpage}
\end{document}